\makeatletter
\def\input@path{{./tex/}{./bst/}}
\makeatother
\documentclass[%
reprint,
superscriptaddress,
 amsmath,amssymb,
 aps,
 prx,
floatfix,
]{revtex4-2}

\usepackage{graphicx}% Include figure files
\usepackage{dcolumn}% Align table columns on decimal point
\usepackage{bm}% bold math
\usepackage{hyperref}% add hypertext capabilities
\usepackage{siunitx}
\usepackage{physics}
\usepackage{upgreek}
\usepackage{comment}

\usepackage[dvipsnames]{xcolor}  % Used to color text during editing
\newcommand{\mose}{MoSe$_2$}
\newcommand{\apone}{AP}
\newcommand{\aptwo}{AP'}

\newcommand{\ct}[1]{\textcolor{green}{\textbf{[cite!]}}}
\newcommand{\abbrev}[3]{
  \newcounter{#3}
  \setcounter{#3}{0}
  \newcommand{#1}{\ifnum\value{#3}<1{#2 (#3)}\else{#3}\fi\stepcounter{#3}}}

\abbrev{\hBN}{hexagonal boron nitride}{h-BN}
\abbrev{\TMD}{transition metal dichalcogenide}{TMD}
\abbrev{\AP}{attractive polaron}{AP}
\abbrev{\RP}{repulsive polaron}{RP}
\abbrev{\MP}{middle polaron}{MP}
\abbrev{\DR}{differential reflection}{DR}
\abbrev{\tBN}{twisted h-BN}{t-BN}

\renewcommand{\pv}{\vb{p}}
\newcommand{\rv}{\vb{r}}
\newcommand{\bv}{\vb{b}}

\newcommand{\Rv}{\vb{R}}
\newcommand{\mv}{\vb{m}}
\newcommand{\mX}{m_{\rm{X}}}
\newcommand{\me}{m_{\rm{e}}}
\newcommand{\rve}{\vb{r}_{\rm e}}
\newcommand{\rveone}{\vb{r}_{\rm{e} 1}}
\newcommand{\rvetwo}{\vb{r}_{\rm{e} 2}}
\newcommand{\rvX}{\vb{r}_{\rm{X}}}
\newcommand{\VH}{V_{\rm{H}}}
\newcommand{\UM}{U_{\rm M}}

\begin{document}

%\preprint{APS/123-QED}

\title{Confined Trions and Mott--Wigner States in a Purely Electrostatic Moiré Potential}

\author{Natasha Kiper}
\email{nkiper@phys.ethz.ch}
\affiliation{Institute for Quantum Electronics, ETH Zürich, Zürich, Switzerland}
\author{Haydn S. Adlong}
\affiliation{Institute for Quantum Electronics, ETH Zürich, Zürich, Switzerland}
\affiliation{Institute for Theoretical Physics, ETH Zürich, Zürich, Switzerland}
\author{Arthur Christianen}
\affiliation{Institute for Quantum Electronics, ETH Zürich, Zürich, Switzerland}
\affiliation{Institute for Theoretical Physics, ETH Zürich, Zürich, Switzerland}
\author{Martin~Kroner}%
\affiliation{Institute for Quantum Electronics, ETH Zürich, Zürich, Switzerland}
\author{Kenji Watanabe}
\affiliation{National Institute for Materials Science, 1-1 Namiki, Tsukuba 305-0044, Japan}
\author{Takashi Taniguchi}
\affiliation{National Institute for Materials Science, 1-1 Namiki, Tsukuba 305-0044, Japan}
\author{Atac {\.I}mamo{\u{g}}lu}
\affiliation{Institute for Quantum Electronics, ETH Zürich, Zürich, Switzerland}

\date{\today}% It is always \today, today,

\begin{abstract}
Moiré heterostructures consisting of \TMD\ hetero- and homobilayers have emerged as a promising material platform to study correlated electronic states.  Optical signatures of strong correlations in the form of Mott--Wigner states and fractional Chern insulators have already been observed in \TMD\ monolayers and their twisted bilayers. In this work, we use a moiré substrate containing a twisted \hBN\ interface to externally generate a superlattice potential for the \TMD\ layer: the periodic structure of ferroelectric domains in \hBN\ effects a purely electrostatic potential for charge carriers.  
We find direct evidence for the induced moiré potential in the emergence of new excitonic resonances at integer fillings, and our observation of  an enhancement of the trion binding energy by $\simeq 3$~meV. A theoretical model for exciton-electron interactions allows us to directly determine the moiré potential modulation of $30 \pm 5$~meV from the measured trion binding energy  shift. We obtain direct evidence for charge order linked to electronic Mott--Wigner states at filling factors $\nu =1/3$ and $\nu=2/3$ through the associated exciton Umklapp resonances.
\end{abstract}
\maketitle

\section{Introduction} \label{sec:introduction}

Semiconductor moiré materials have emerged as a fertile platform to explore the physics of strongly correlated electrons \cite{reganMottGeneralizedWigner2020,tangSimulationHubbardModel2020, zhangTwistangleDependenceMoire2020,ciorciaroKineticMagnetismTriangular2023}. Generically,  transition metal dichalcogenide (TMD) moiré materials are realized by interfacing two or more monolayers,  and the moiré pattern emerges due to twist angle or lattice mismatch between the layers. The emerging periodic potential has substantial contribution from lattice reconstruction, in addition to polarization fields and inter-layer tunneling. In these structures, not only charge carriers but also excitons are
typically subject to strong potentials. This feature has been utilized to investigate correlated Mott states of Bose--Fermi mixtures consisting of moiré inter-layer excitons and electrons in heterobilayers with type-II band-alignment \cite{xiongCorrelatedInsulatorExcitons2023, parkDipoleLaddersLarge2023, gaoExcitonicMottInsulator2024}. 

Due to the complex nature of the moiré potential and its different effect on charge carriers and excitons, the interpretation of optical signatures from TMD moiré structures is often challenging. Thus, designing systems where excitons are not subject to a moiré potential would be advantageous, since any modifications of the optical excitation spectrum would arise exclusively from electronic correlations and their effect on exciton-electron interactions. This scenario would be ensured if the moiré potential is of purely electrostatic origin. Moreover, exposing two identical TMD monolayers  that are separated by a thin tunneling barrier to an identical periodic electrostatic potential would lead to moiré electrons or holes acquiring a layer pseudo-spin degree of freedom with (nearly) SU(2) symmetry. A range of fascinating phenomena ranging from electrically controlled Feshbach resonances~\cite{schwartzElectricallyTunableFeshbach2021} to chiral layer pseudo-spin liquids \cite{kuhlenkampChiralPseudospinLiquids2024} could be realized using such a structure.

Here, we demonstrate a purely electrostatic moiré potential for itinerant electrons and holes in a \mose\ monolayer, which we probe through \DR\ spectroscopy. The simple nature of the moiré potential allows us to extract its modulation depth directly from the optical signatures -- namely the trion binding energy -- using a theoretical model for exciton-electron interactions.  In addition to measuring a clear red-shift of the trion and the associated \AP\ resonance, we identify the charge ordered Mott--Wigner states at filling factors $1/3$ and $2/3$ through the appearance of excitonic Umklapp resonances~\cite{shimazakiOpticalSignaturesPeriodic2021}.

\begin{figure}[h]
    \includegraphics[width=\columnwidth]{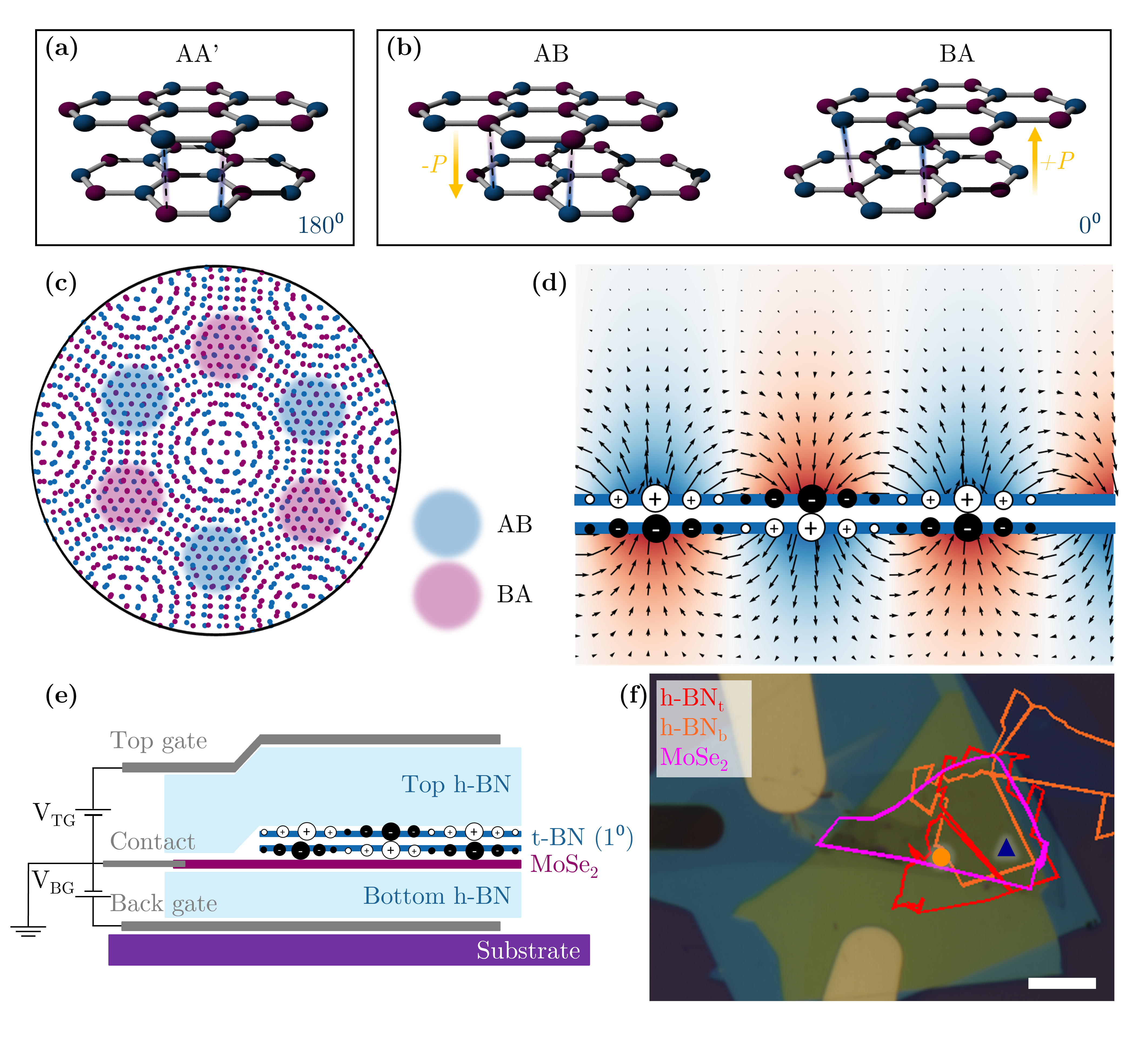}
    \caption{(a) Crystal structure of bulk h-BN (AA'-stacking), corresponding to a 180$^\circ$ rotation between adjacent layers. Each boron atom (blue) is vertically neighbored by a nitrogen atom (purple), and the crystal has an inversion center at the center of the unit cell. (b) The two possible lowest-energy stacking configurations of a 0$^\circ$-aligned h-BN bilayer. In AB-stacking, the nitrogen atoms of the top layer are vertically aligned to boron atoms in the bottom layer, and vice versa in BA-stacking. The crystal no longer has inversion symmetry and a spontaneous out-of-plane polarization develops. The polarization has equal magnitude but opposite direction for the two stacking configurations. (c) Schematic of moiré pattern formed by two h-BN layers with a twist angle close to 0$^\circ$. It contains alternating regions of AB- and BA-stacking. (d) Side view of a twisted h-BN bilayer with alternating electric dipoles in adjacent AB- and BA-stacked regions. The electric field and potential generated around the bilayer are shown as arrows and red-blue coloring, respectively. (e) Schematic of the device structure and electrical connections. The twisted h-BN interface was fabricated from a single h-BN monolayer using the tear-and-stack technique. (f) Optical micrograph of the device with flake outlines. The orange circle and blue triangle indicate the measurement spots of spectra shown in Fig.~\ref{fig:2}(a) and (b), respectively. Scale bar:~10 $\upmu$m.
    }
    \label{fig:1}
\end{figure}

\section{Electrostatic moiré potential generated by twisted hexagonal boron nitride} \label{sec:hbn}

We generate a purely electrostatic periodic potential using ferroelectric domains in twisted hexagonal boron nitride (\hBN), which is a layered insulator with a large band gap. In bulk, \hBN\ naturally occurs with AA’ stacking, where neighboring layers are rotated by 180$^\circ$ with respect to each other [see Fig.~\ref{fig:1}(a)]: inversion symmetry of the structure ensures a weak paraelectric response to applied out-of-plane fields. In AB-stacked or 0$^\circ$ aligned \hBN, however, the two neighboring layers have the same orientation. The lattice can now take one of two possible stacking orders: AB and BA stacking. Both break inversion and out-of-plane mirror symmetry [see Fig.~\ref{fig:1}(b)]. Therefore, an out-of-plane electric polarization develops spontaneously and the bilayer becomes ferroelectric \cite{gilbertAlternativeStackingSequences2019, kimStackingOrderDependent2013, zhouVanWaalsBilayer2015, constantinescuStackingBulkBilayer2013, yasudaStackingengineeredFerroelectricityBilayer2021}. A small twist angle between the \hBN\ layers leads to a spatial modulation of the local stacking order, resulting in alternating AB and BA domains with opposite out-of-plane electrical polarization [see Fig.~\ref{fig:1}(c)]. These alternating electric dipoles create electric fields that extend beyond the twisted bilayer and effect a superlattice potential for charges in any target material placed close to the twisted interface [see Fig.~\ref{fig:1}(d)] \cite{zhaoUniversalSuperlatticePotential2021, woodsChargepolarizedInterfacialSuperlattices2021, kimElectrostaticMoirePotential2024a, wangBandStructureEngineering2024, zhangEmergenceMoireSuperlattice2024}.

We place a \mose\ monolayer directly below a \tBN\ bilayer in a dual-gated, charge tunable van der Waals heterostructure [Fig.~\ref{fig:1}(e)-(f)]. The purely electrostatic nature of the moiré potential for electrons and holes in \mose\  is ensured by two key features of the \hBN\ \TMD\ interface: namely, (i) the lattice mismatch exceeding $20 \%$, and (ii)  the large difference between the conduction (valence) band minimum (maximum) of the two materials. The sizeable lattice mismatch ensures that no large scale moiré pattern is formed between the \hBN\ and \TMD\ layers directly, while the band alignment ensures that charges are injected solely into the \mose\ layer. Based on these features, we assume the electronic moiré potential to have a simple shape determined purely by the local stacking order of the \tBN\ layers, with a single minimum per moiré unit cell and no secondary local minima. Up to unity filling of the moiré lattice, we expect charged particles to populate the moiré flat band originating from hopping between these minima. We also note that we expect electrons and holes to experience potentials of equal shape and depth, but of opposite sign. These features render \tBN-generated moiré potentials an ideal platform to investigate strong electronic correlations in \TMD\ moiré materials.

\section{Basic Characterization} \label{sec:basics}

We use \DR\  measurements of the \mose\ monolayer to determine the strength of the periodic potential experienced by charge carriers through the modification of the trion binding energy. We designed our sample to ensure that the  \mose\ layer extends beyond the \tBN\ region, allowing us to directly compare the experimental signatures with (region I) and without (region II) moiré potential in a single device (see Appendix~\ref{sec:app_fab} for details on sample fabrication). In the charge neutral regime, the DR spectrum is indistinguishable between regions I and II, verifying the absence of a moiré potential for excitons. This is in stark contrast to conventional hetero-bilayer TMD moiré systems \cite{jinObservationMoireExcitons2019, tranEvidenceMoireExcitons2019, seylerSignaturesMoiretrappedValley2019, zhangVanWaalsHeterostructure2021, campbellExcitonpolaronsPresenceStrongly2022, ciorciaroKineticMagnetismTriangular2023, polovnikovImplementationBilayerHubbard2024, zhaoHybridMoireExcitons2024}, where the presence of a moiré superlattice leads to dramatic modifications of the exciton spectrum. 

\begin{figure}[h]
    \includegraphics[width=\columnwidth]{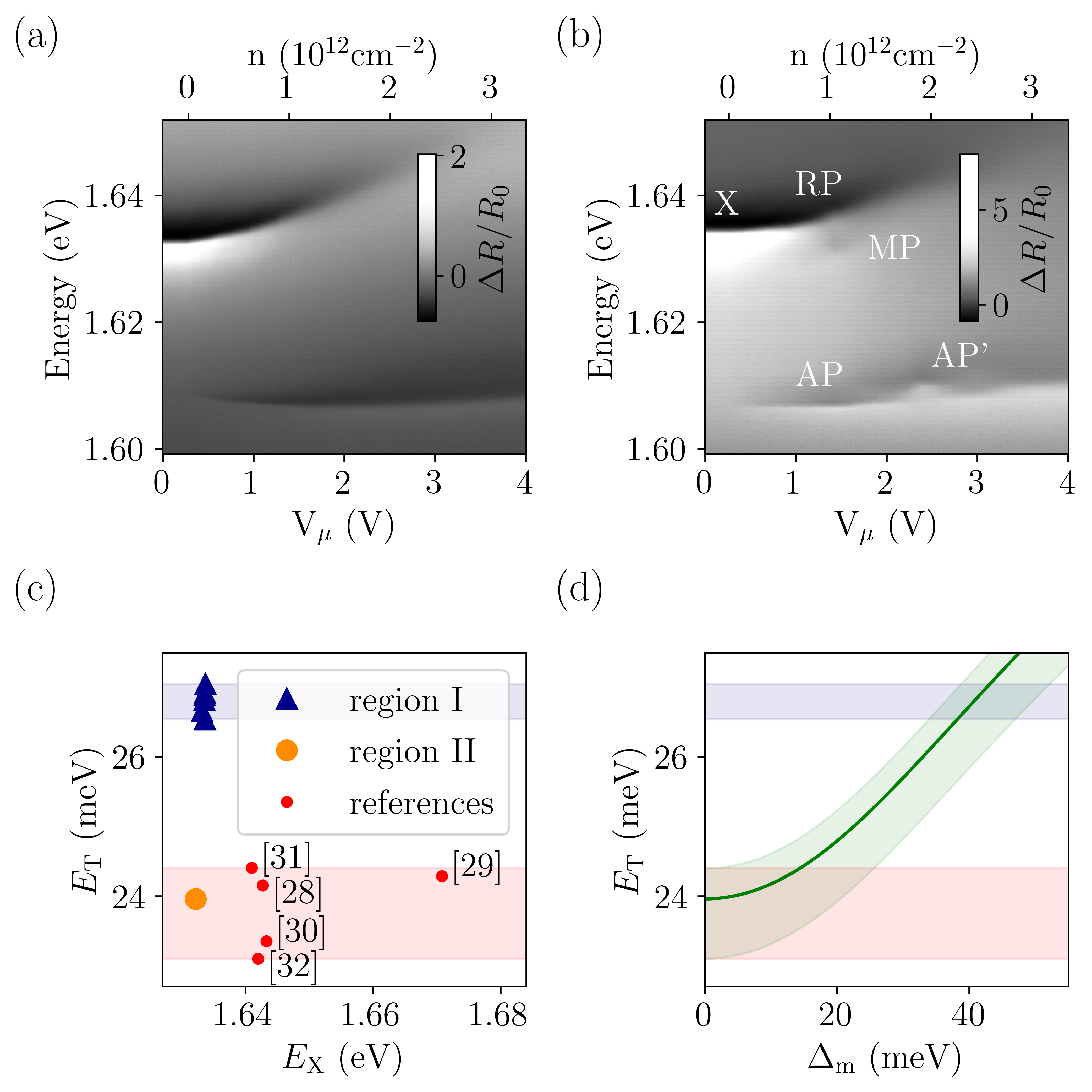}
    \caption{(a) Differential reflection spectrum of moiré-free \mose\ in region II as a function of $V_\mu=0.42V_\text{TG}+0.58V_\text{BG}$. In charge neutrality, there is one strong exciton resonance, which splits into attractive and repulsive polarons upon doping with electrons. (b) Differential reflection spectrum as a function of $V_\mu$ in region I. The spectrum at charge neutrality is qualitatively unchanged, but new resonances appear at certain doping levels. (c) Comparison of trion binding energies for moiré-confined and free trions. Blue triangles mark trion binding energy $E_\text{T}$ vs. neutral exciton energy $E_\text{X}$ on six different spots in region I. Orange circle shows $E_\text{T}$ vs. $E_\text{X}$ for the spectrum shown in panel (a), red markers show $E_\text{T}$ vs. $E_\text{X}$ values measured on five different h-BN-encapsulated monolayer \mose\ devices \cite{smolenskiInteractionInducedShubnikovHaas2019, ciorciaroObservationMagneticProximity2020, thurejaElectricallyTunableQuantum2022, thurejaElectricallyDefinedQuantum2024,liuExcitonpolaronRydbergStates2021}. The light blue and light red bands indicate the spread of binding energies for the moiré-bound and free trions. (d) The theoretically calculated $E_\text{T}$ as a function of moiré depth $\Delta_\text{m}$. The light blue and light red bands show the same $E_\text{T}$ intervals as in (c). The shaded area around the curve shows the variation corresponding to the spread of measured $E_\text{T}$ in the absence of moiré. 
    }
    \label{fig:2}
\end{figure}

Figures~\ref{fig:2}(a)-(b) compare the \DR\ spectrum  of the \mose\ monolayer with and without an electrostatic moiré potential as a function of the chemical potential $V_\mu=0.42V_\text{TG}+0.58V_\text{BG}$: \DR\ depicted in (a) and (b) is obtained at a spot in region II and I, respectively. The two measurement spots are indicated by the orange circle and blue triangle in Fig.~\ref{fig:1}(f). We calculate the charge density corresponding to the applied gate voltage using the capacitor model, and using the integrated DR around the exciton resonance energy to determine the doping onset (see Appendix~\ref{sec:app_ncalc} for details). We keep the applied out-of-plane displacement field $V_E=0.5(V_\text{TG}-V_\text{BG})$ constant at $V_E=0$. 
In the charge neutral regime ($0\leq V_\mu\leq 0.26$~V), there is a single resonance in the spectrum at $1.632$~eV (a) and $1.634$~eV (b), corresponding to the A exciton of \mose; the small ($2$~meV) difference in the resonance energy between (a) and (b) is likely to stem from local strain variation, which is known to result in a shift of the neutral exciton energy by as much as $5$~meV in \hBN-encapsulated \mose\ monolayers free of external potentials. 

The absence of additional excitonic resonances and lack of reduction in the DR contrast of the A exciton in Fig.~\ref{fig:2}(b) verify that excitons in our sample are not subject to a significant moiré potential. We emphasize that while electrons and holes experience a moiré potential whose strength is linearly proportional to the out-of-plane electric field ($E_\text{z,moiré}$) generated by the ferroelectric domains, we would expect charge-neutral excitons to experience a much weaker moiré potential that originates from the dc-Stark shift with a strength proportional to the square of the in-plane electric field \cite{thurejaElectricallyTunableQuantum2022}. Since we do not observe any additional exciton resonances in charge neutrality, we conclude that the energy splitting arising from dc Stark effect is smaller than the exciton linewidth $\gamma=2.3$~meV. Therefore, we estimate a maximal in-plane field $|E_{||,\text{max}}|=25$~mV/nm, assuming an exciton polarizability $\alpha=6.5$~eV\,nm$^2$\,V$^{-2}$ \cite{cavalcanteStarkShiftExcitons2018}. 

Upon doping the monolayer in region II with electrons for $V_\mu\geq0.26$~V, the \AP\ resonance appears, while the exciton evolves into a \RP\ which shifts to higher energies and broadens with increasing electron density [Fig.~\ref{fig:2}(a)]. In the limit of vanishing doping, the \AP\ is separated from the neutral exciton by the trion binding energy $E_\text{T}=24$~meV.  The \AP\ gains oscillator strength with increased electron doping and exhibits first a slight redshift up to $V_\mu\approx 1.5$~V followed by an overall blueshift. 

\section{Extraction of moiré depth from increased trion binding energy} \label{sec:depth}

The \AP\ in region I [Fig.~\ref{fig:2}(b)] shows striking differences as compared to the electron-doped monolayer in region II [Fig.~\ref{fig:2}(a)]: by focusing on the low electron doping regime, we find that the trion binding energy is increased by $2.75$ meV to $E_\text{T,moiré}=26.75$ meV.  
Figure~\ref{fig:2}(c) shows trion binding energies versus exciton energies on six different \tBN-adjacent spots in region I of our device, as well as one  spot in region II. Additionally, we show trion binding energies on five different simple \hBN-encapsulated \mose\ monolayer devices for comparison. Although the A exciton resonance energies vary by more than $10$~meV between different devices due to different local strain and dielectric environments, the moiré-free trion binding energy is very uniform (within $\sim1$~meV). In stark contrast, $E_{\text{T,moiré}}$ is clearly enhanced on every \tBN-adjacent spot measured on this sample. This increase in  $E_\text{T,moiré}$ in comparison to $E_\text{T}$ is remarkable, given the uniformity of the trion binding energy despite the exciton energy variation.

The increased trion binding energy $E_{\text{T,moiré}}$ is a direct consequence of the localization of the electron, and consequently of the trion, wave function in the superlattice potential. Intuitively, since the electron is already confined by the moiré potential, the kinetic energy cost of correlating its motion with the exciton is reduced as compared to the free electron case. To verify this explanation, we carried out a calculation quantifying the modification of the trion binding energy in the presence of a moiré potential for the electron (see Appendix~\ref{sec:app_trionbinding}). Assuming that the exciton Bohr radius is much smaller than all other relevant length scales, we use an effective short range exciton-electron interaction potential that correctly describes the binding energy of the trion in the absence of the moiré potential. We find that the $2.75$~meV increase in trion binding energy is consistent with a moiré potential modulation ($\Delta_\text{m}$) in the range $30$~meV $\lesssim \Delta_\text{m} \lesssim 40$~meV, assuming a moiré lattice constant of $a_\text{m} = 10.2$~nm [see Fig.~\ref{fig:2}(d) and Fig.~\ref{fig:si_theory}]. The main contribution to the uncertainty in the potential depth is the spread of measured trion binding energies. The size of the electron Wannier orbital for such a moiré potential is $a_\text{W} \simeq 3.6$~nm~\footnote{We estimate the size of the Wannier orbital as the variance of the numerically calculated electron Wannier function.}, which is comparable to but larger than the trion Bohr radius $a_\text{T} \simeq 2.1$~nm. We remark that the calculated $E_\text{z,moiré}$ based on previous reports \cite{zhaoUniversalSuperlatticePotential2021} predicts a factor of $4$ larger $\Delta_\text{m}$ value as compared to what we determine from our measurements. We tentatively attribute the difference to the significant dielectric screening of $E_\text{z,moiré}$ by the \mose\ monolayer. This value of $\Delta_\text{m}$ combined with a domain length scale of $l\sim 10$~nm is also consistent with the absence of splitting of the exciton due to the dc Stark effect, as $|E_{||}| \sim\Delta_\text{m}/l\ll |E_{||,\text{max}}|=25$~mV/nm. 

\begin{figure}[h]
    \centering
    \includegraphics[width=\columnwidth]{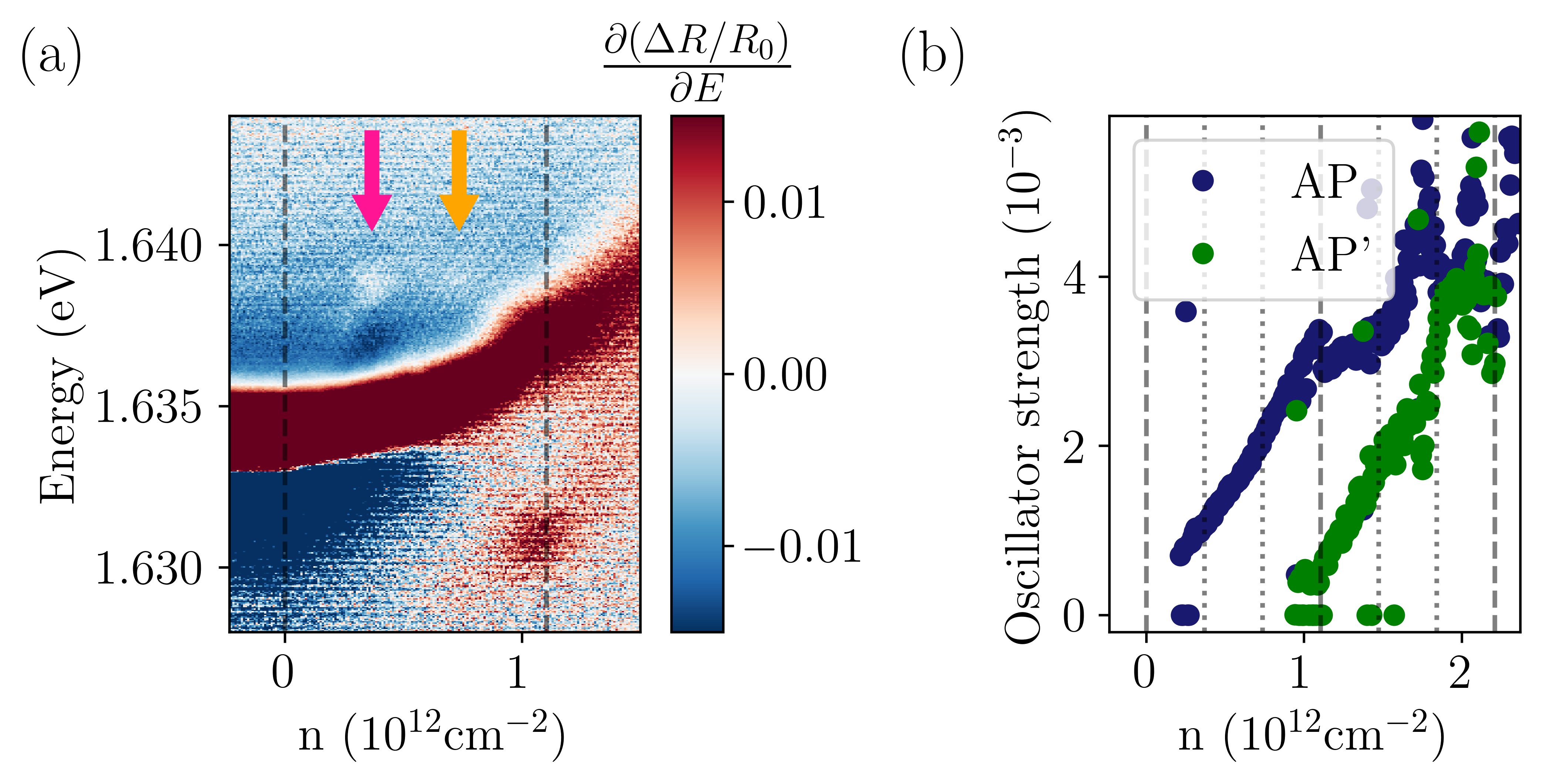}
    \caption{(a) Differentiated DR spectrum of the repulsive polaron as a function of electron density at low doping. the pink and orange arrows indicate the Umklapp resonances at $\nu=1/3$ and $\nu=2/3$, respectively. The gray dashed lines show the onset of doping and the doping level at filling factor $\nu=1$ (one electron per moiré unit cell). (b) Oscillator strengths of the \apone\ and \aptwo\ resonances obtained from fitting the spectra in Fig.~\ref{fig:2}(b) with dispersive Lorentzian functions. The dashed vertical lines correspond to integer fillings, the dotted lines to $\nu=1/3,2/3,4/3,$ and $5/3$.}
    \label{fig:3}
\end{figure}

\section{Mott--Wigner states at fractional fillings of the moiré lattice} \label{sec:umklapp}

At $n\approx 3.7\cdot10^{11}$~cm$^{-2}$ and $n\approx7.4\cdot10^{11}$~cm$^{-2}$, two new, much weaker resonances appear at slightly higher energy than the \RP\ [see Fig.~\ref{fig:3}(a)]. We associate them with Umklapp resonances arising from the scattering of the \RP\ from the periodic potential that arises from electronic charge order \cite{shimazakiOpticalSignaturesPeriodic2021,smolenskiSignaturesWignerCrystal2021}. The sudden appearance and disappearance of the Umklapp peaks indicates that the electrons form a Mott--Wigner state at these specific densities. Based on previous reports of insulating states at fractional fillings of TMD moiré lattices \cite{xuCorrelatedInsulatingStates2020}, we assign them to be the $\nu=1/3$ and $\nu=2/3$ states, as these are usually the most robust fractionally filled states. Although the electron density at $\nu=2/3$ is twice that at $\nu=1/3$,
the two Umklapp resonances appear approximately at the same blue detuning. This confirms the expectation that at $\nu=2/3$, electrons form a hexagonal lattice with the same unit cell size as the triangular lattice they form at $\nu=1/3$, and is in agreement with prior theoretical predictions~\cite{salvadorOpticalSignaturesPeriodic2022} (see also Appendix~\ref{sec:app_umklapp}). The Umklapp resonances allow us to accurately determine the electron density corresponding to integer fillings. We find $n|_{\nu=1}=(1.1\pm0.2)\cdot10^{12}$~cm$^{-2}$, yielding a moiré lattice length $a_\text{m}=10.2\pm0.7$~nm. The corresponding \tBN\ twist angle of $\theta_\text{m,exp} = 1.4\pm0.1^\circ$ is in relatively good agreement with the angle $\theta_\text{m,tar}=1^\circ$ we targeted during the fabrication of the heterostructure. The onset of doping and $n|_{\nu=1}$ are indicated by the dashed gray lines in Fig.~\ref{fig:3}(a).

\section{Characterization of the moiré-induced polaron resonances} \label{sec:ap}

The \DR\ spectrum in Fig.~\ref{fig:2}(b) shows that there are two new resonances that appear upon increasing the electron density, which we term the {\sl second attractive polaron} (\aptwo) and the {\sl middle polaron} (MP). The MP attains its maximum strength at $n|_{\nu=1} \simeq 1 \cdot10^{12}$~cm$^{-2}$, where we also see the onset of \aptwo, strongly indicating that they are linked to the electronic Mott state at $\nu=1$. We emphasize that we observe the same new spectral features -- the \aptwo\ and MP resonances -- when the monolayer is hole doped (see Appendix~\ref{sec:app_holeside}).

We extract the oscillator strengths of the \apone\ and \aptwo\ resonances by fitting the spectra with dispersive Lorentzian functions [see Fig.~\ref{fig:3}(b) and Appendix~\ref{sec:app_analysis}]. We find that the oscillator strength of the \apone\ resonance increases linearly up to $\nu=1$, where it saturates.  Meanwhile, the \aptwo\ resonance  increases linearly with $n$ between $\nu=1$ and $\nu=5/3$. Interestingly, both \AP\ resonances' oscillator strengths exhibit nonlinear behaviors at $\nu=5/3$, hinting at the presence of correlated states at higher fractional fillings that are not detectable by Umklapp spectroscopy.

\begin{figure}[h]
    \includegraphics[width=\columnwidth]{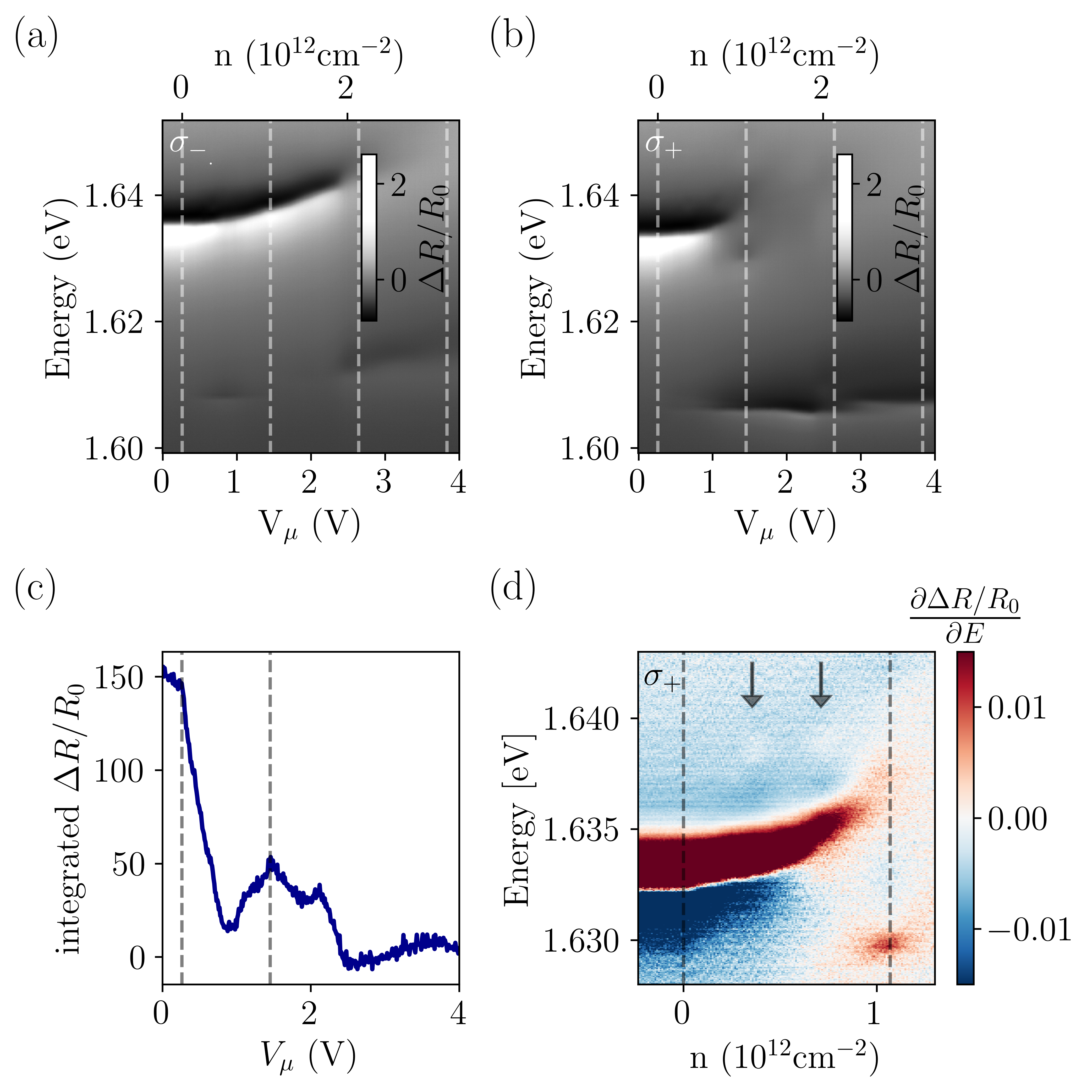}
    \caption{(a) and (b) Left- and right-hand circular polarized DR spectra at $B_\text{z}=7$~T. Dashed vertical lines indicate integer filling factors of the moiré lattice. (c) Integrated left-hand circular polarized DR spectra for energies between 1.634~eV and 1.652~eV as a function of $V_\mu$. We use the abrupt drop at $V_\mu\approx0.3$~V and the local maximum at $V_\mu\approx1.5$~V to determine the doping onset and unity filling ($\nu=1$, or one electron per moiré unit cell). (d) Differentiated right-hand circular polarized DR spectra, showing the Umklapp resonances at $\nu = 1/3$ and $\nu = 2/3$. The dashed gray lines correspond to the doping onset and $\nu = 1$.}
    \label{fig:4}
\end{figure}

To better understand the new features induced by the \tBN-moiré potential, we apply a magnetic field of $B_\text{z}=7$~T in the out-of-plane direction. The optical selection rules and spin-valley locking due to spin-orbit coupling in \mose\ make it possible to optically probe the density of spin-up and spin-down electrons using circular polarization-resolved DR spectroscopy. Since a bound trion state composed of two electrons and one hole is only formed when the two bound electrons have a symmetric orbital wavefunction centered around the hole, only electrons occupying K (-K) valley can dress a {-K} (K) valley exciton to form an \AP\ \cite{sidlerFermiPolaronpolaritonsChargetunable2017,efimkinManybodyTheoryTrion2017}. Figures~\ref{fig:4}(a) and (b) show \DR\ spectra for K ($\sigma_+$) and -K ($\sigma_-$) valley exciton-polaron resonances at $B_\text{z} = 7$~T.

For $\sigma_+$-polarization [Fig.~\ref{fig:4}(b)], we see an \AP\ originating from a trion with binding energy of $E_\text{T}=26.6$~meV; the measured $E_\text{T}$ is identical to the one measured at $B_\text{z}=0$~T, confirming that the confinement from the \tBN\ electrostatic moiré potential in the limit of vanishing electron density is unaffected by the external magnetic field. 

The $\sigma_-$-polarized exciton cannot form a bound state with the electrons, but experiences a blueshift due to phase space filling and broadening due to intravalley electron scattering. We observe a cusp-like feature in the energy and a local maximum in integrated DR spectrum of the $\sigma_-$ exciton at $V_\mu=1.45$~V [see Fig.~\ref{fig:4}(a) and (c)]: such features have been previously reported for excitonic excitations out of incompressible electronic states \cite{smolenskiInteractionInducedShubnikovHaas2019}. Assuming that this cusp is associated with the $\nu=1$ Mott state, we find a twist angle $\theta_\text{m}^\text{exp}=1.36\pm0.09^\circ$, which in turn is (within the measurement uncertainty) identical to the one we determined independently at $B_\text{z}=0$~T. We also observe the Umklapp features at $B_\text{z} = 7$~T [see Fig.~\ref{fig:4}(d) and Fig.~\ref{fig:si_umklapp}]. 

The absence of an \AP\ resonance in $\sigma_-$-polarization for $n\leq1.9\cdot10^{12}$~cm$^{-2}$~\footnote{We attribute the weak AP resonance in $\sigma_-$-polarization we observe for $n \le 0.4\times10^{12}$~cm$^{-2}$ to probe-light-induced spin-valley depolarization~\cite{smolenskiSpinValleyRelaxationExcitonInduced2022,ciorciaroKineticMagnetismTriangular2023}.}
indicates that, similarly to the moiré-free case \cite{backGiantParamagnetismInducedValley2017}, the electrons at these densities are completely spin-polarized due to the valley-Zeeman effect \cite{srivastavaValleyZeemanEffect2015}. This observation in turn indicates that for $1.0 < \nu < 1.8$  the electrons in doubly occupied sites are in the $m_\text{z} = 1$ triplet ($T_+$) state. The fact that we can polarize the spins completely even above $\nu=1$ implies that the valley-Zeeman energy is larger than the singlet-triplet energy splitting in doubly occupied sites. This conclusion is further supported by exact Diagonalization (ED) calculations for two electrons confined to a single moiré unit cell, where we account for neighboring electrons using a Hartree potential (see Appendix~\ref{sec:app_theorynu2}). In particular, owing to the strong Coulomb interactions, we find that the triplet exceeds the singlet in energy by only 0.38~meV at $\nu=2$. 

The AP' resonance disappears when the electrons are valley polarized at $B_\text{z} = 7$~T and when $\nu<1$ at $B_\text{z} = 0$~T. This suggests that AP' is the optical excitation out of doubly occupied moir\'e sites where the electrons form a singlet or $m_\text{z}=0$ triplet (T$_0$) state, measured with respect to the quantization axis set by the spin of the exciton (see Appendix~\ref{sec:app_theorynu2} for details). 
On these sites, only one electron can participate in trion formation, and the other electron leads to a blueshift of the polaron resonance due to phase space filling. The energy difference between the AP and AP' resonances is approximately reproduced by a phenomenological model at $\nu=2$, based on ED calculations of an exciton and two electrons in a single unit cell. We provide a detailed discussion of the theoretical model in Appendix~\ref{sec:app_theorynu2}.

The relative AP and AP' spectral weights for a given exciton spin$/$polarization can be understood as follows. For $B_\text{z}= 0$~T and $1 < \nu < 2$, 50\% of the doubly occupied sites are either in singlet or T$_0$ states, and yield the blue shifted AP' resonance. The singly occupied sites and doubly occupied T$_+$ sites give rise to AP. T$_+$ sites contribute doubly to the oscillator strength, since the exciton can bind to either electron. On T$_-$ sites, no attractive polarons or trions can be formed.
At $B_\text{z}= 7$~T, electrons in all doubly occupied sites are in T$_+$ states (for $\sigma_+$-polarization), and we only observe the AP resonance.

The MP resonance also appears only in $\sigma_+$ polarization, indicating that it originates from excitons interacting with opposite valley electrons. Given the experimental signatures at $B_\text{z}=0$~T and $B_\text{z}=7$~T, we expect that the MP resonance arises from the attractive Hartree potential that excitons are subject to when the electrons form a Mott state at  $\nu \simeq 1$. Theoretical modeling of this Hartree potential indeed gives rise to a new negative-energy excitonic band, and the large oscillator strength observed experimentally is consistent with delocalized excitons. What remains to be better understood is the character of MP at $B_{\rm z} =0$ T, where we expect that the effective Hartree potential fluctuates sizeably from site to site due to the strong dependence of the exciton-electron interaction on the electron spin.

\section{Conclusion} \label{sec:conclusion}

We have presented unequivocal evidence for the realization of a purely electrostatic moiré potential for charge carriers, and the formation of generalized Wigner states. Using the enhanced trion binding energy, we have determined the moiré potential generated by twisted \hBN\ is at least a factor of two weaker than those reported in TMD heterobilayers~\cite{jinObservationMoireExcitons2019, shabaniDeepMoirePotentials2021, ciorciaroKineticMagnetismTriangular2023}. This system hosts a rich interplay between electronic Coulomb interactions and the moiré potential, revealing qualitatively new features such as the second attractive polaron and the middle polaron. Owing to the electrostatic nature of the moiré potential, we can use the charge-neutral excitons to gain valuable insights into these features, such as finding strong evidence for the origin of AP' in the spin-structure of doubly occupied sites in the Hartree-moiré potential. 

This new platform for moiré provides a number of interesting avenues for future work, such as placing two TMD layers above a \tBN\ interface, giving the electrons an additional layer pseudo-spin degree of freedom with weakly broken SU(2) symmetry. This structure could be used to explore a number of exotic phenomena, ranging from spin-polaron formation through kinetic pairing \cite{prichardDirectlyImagingSpin2024,moreraAttractionKineticFrustration2024}, to chiral layer-pseudospin liquids \cite{zhangSUChiralSpin2021, kuhlenkampChiralPseudospinLiquids2024}.

The data that supports the findings of this paper is available in the ETH Research Collection \cite{ethResearchCollection}.

{\sl Acknowledgements} This work was supported by the Swiss National Science Foundation (SNSF) under Grant Number 200021-204076.
H.S.A acknowledges support from the Swiss Government Excellence Scholarship. A.C. was supported by an ETH Fellowship.
K.W. and T.T. acknowledge support from the JSPS KAKENHI (Grant Numbers 20H00354, 21H05233 and 23H02052) and World Premier
International Research Center Initiative (WPI), MEXT, Japan. We thank Tomasz Smolenski, Ivan Morera and Livio Ciorciaro for useful discussions.

\appendix

\section{Sample Fabrication} \label{sec:app_fab}

\begin{figure}[h]
    \centering
    \includegraphics[width = \columnwidth]{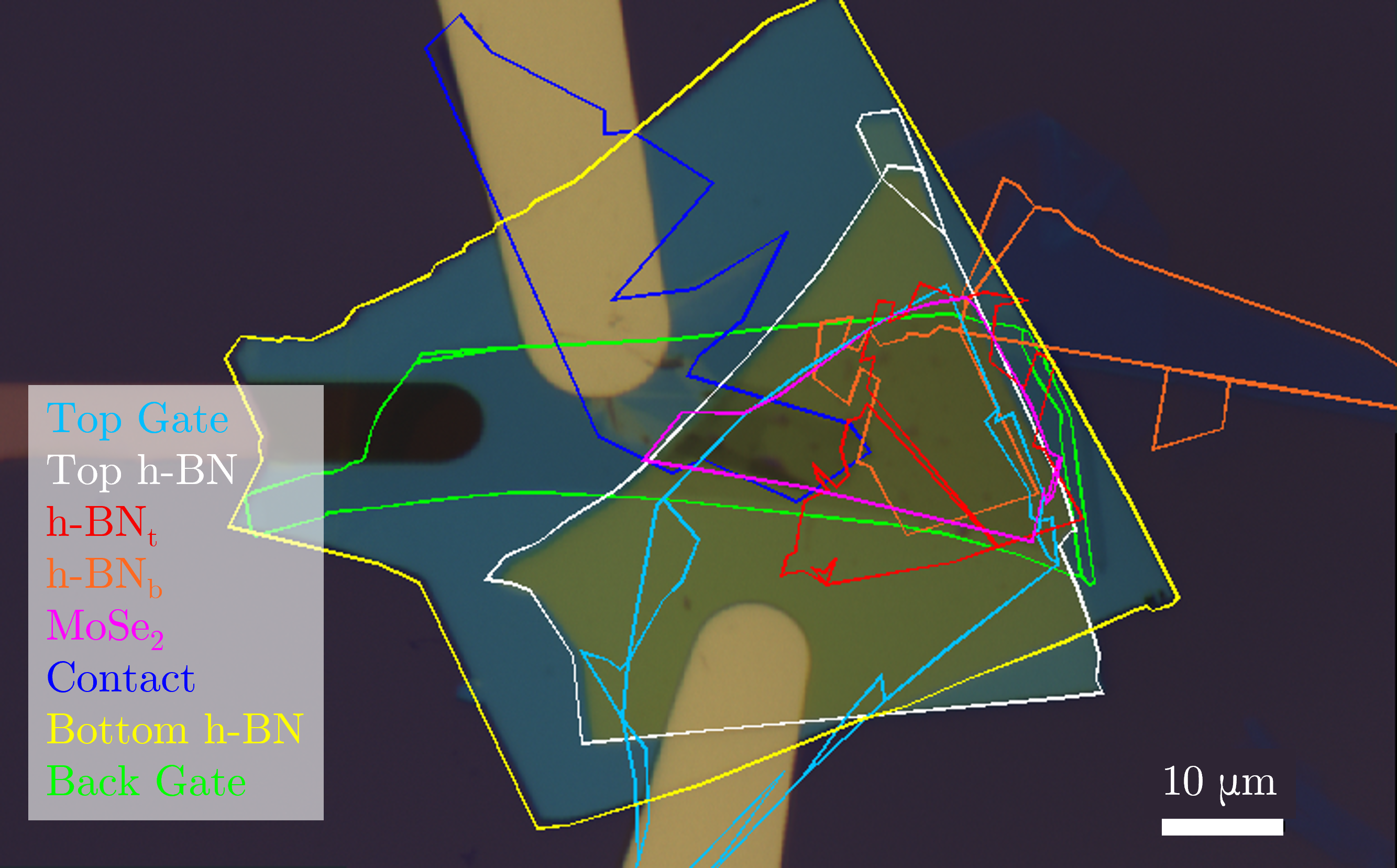}
    \caption{Optical micrograph of the device, including outlines of all flakes.}
    \label{fig:si_sample}
\end{figure}

All flakes were mechanically exfoliated from bulk crystals (\mose\ from HQ Graphene, \hBN\ from our collaborators at NIMS, natural graphite) on Si substrates with a 285~nm SiO$_2$ capping layer. The thicknesses of all flakes were determined by their optical contrast in bright field microscopy. The monolayer nature of the \hBN\ that was used to make the twisted interface was further confirmed by measuring its second harmonic generation. The heterostructure was assembled with a standard dry transfer technique using a poly(bisphenol A carbonate) (PC) covered hemispherical polydimethylsiloxane (PDMS) stamp and deposited on a Si/SiO$_2$ (285~nm) substrate. The stacking was done at high temperature (140~$^\circ$C) inside a glove box in inert Ar atmosphere. The PC was delaminated from the PDMS by heating it to 180~$^\circ$C and subsequently dissolved in chloroform. The twisted \hBN\ was fabricated from a single \hBN\ monolayer using the tear-and-stack technique. Metallic contacts (5~nm Ti sticking layer, 85~nm Au) to the gate and contact graphite flakes were prepared using optical lithography and electron beam evaporation. See Fig.~\ref{fig:si_sample} for the detailed layout of the device.

\section{Experimental Setup} \label{sec:app_setup}

\begin{figure}[h]
    \centering
    \includegraphics[width = \columnwidth]{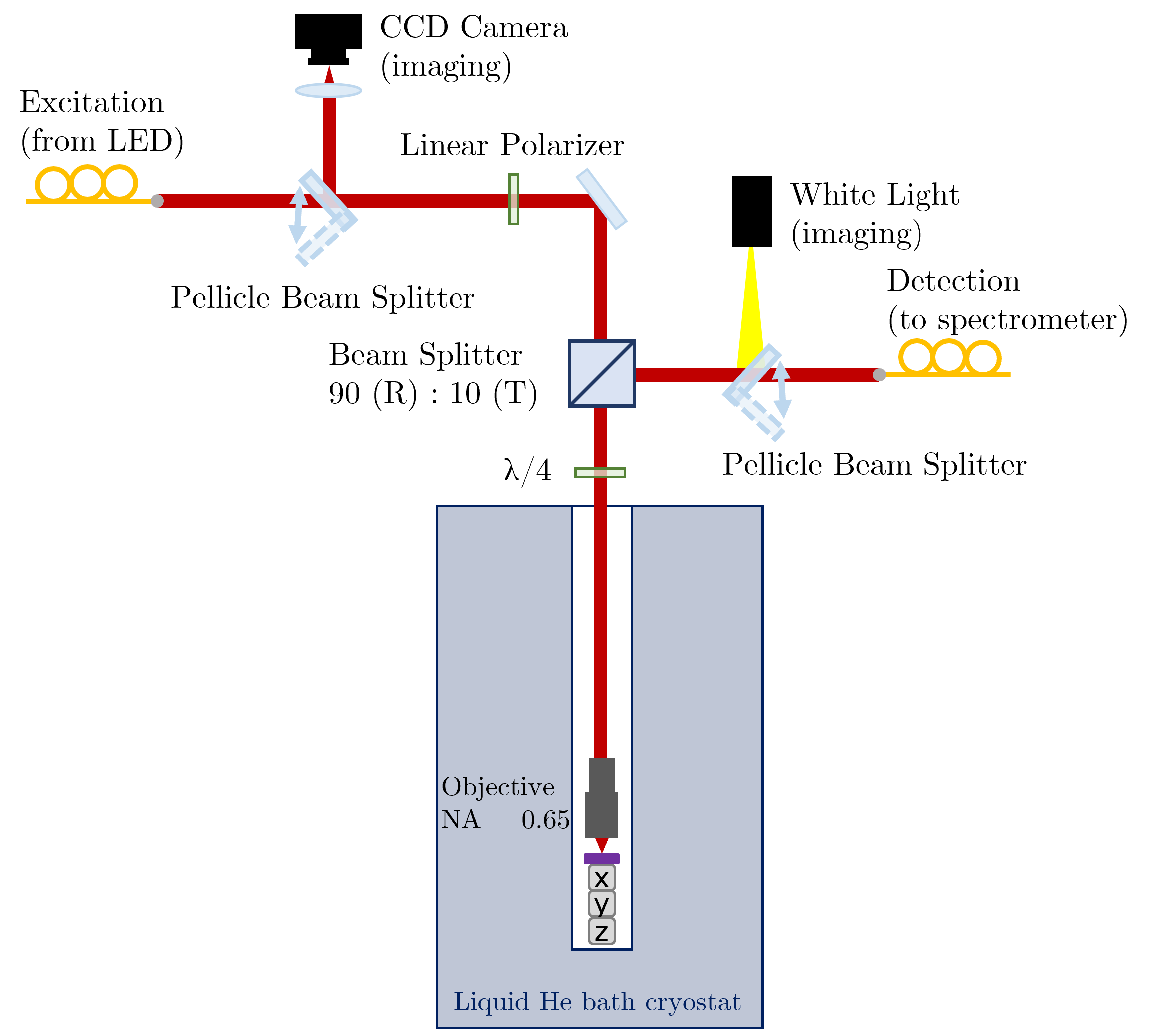}
    \caption{Schematic representation of the measurement setup used for DR measurements.}
    \label{fig:si_setup}
\end{figure}

All DR measurements were carried out using a confocal microscope setup with the sample at $T = 4.2$~K. The entire setup is schematically depicted in Fig.~\ref{fig:si_setup}. The sample was mounted on top of $x-y-z$ piezoelectric positioners at the bottom of a stainless steel tube containing $20$~mbar He exchange gas, which was submerged in a liquid He bath cryostat equipped with a superconducting magnet coil. The light source for DR measurements was a single-mode fiber-coupled broadband ($3$~dB bandwidth of $20$~nm) light emitting diode (LED) centered at $760$~nm ($1.631$~eV), from Exalos. After exiting the fiber, the excitation light was collimated before entering the cryostat insert through a window at the top. The light was then focused on the sample surface by a long working distance apochromatic cryogenic objective (LT-APO/LWD/VISIR/0.65, NA$=0.65$). The reflected light from the sample was collected by the same objective and separated from the excitation light by a 90 (R) : 10 (T) beam splitter. It was then coupled into a single mode fiber which brought it to a $0.75$~m spectrometer equipped with a liquid nitrogen cooled CCD camera and a $1200$ groove/mm diffraction grating.
Two pellicle beam splitters on flip mounts allowed us to selectively inject white light for imaging into the detection path, and image the sample using a CCD camera in the excitation path. 
For the circular-polarization resolved DR measurements at $B_\text{z}=7$~T, we set the polarization to be parallel to the axis of the beam splitter using a linear polarizer in the excitation path, before converting it to a circular polarization using a quarter-wave ($\lambda/4$) plate in the common excitation on detection path. The $\lambda/4$ plate was mounted on a motorized rotation stage, allowing us to switch between $\sigma_+$ and $\sigma_-$ polarization. The circular polarization was set by suppressing the reflection of the excitation light from the sample on the imaging CCD.
For all DR measurements, the excitation power was kept below $10$~nW. 

\section{Analysis of Differential Reflection Spectra} \label{sec:app_analysis}

To obtain the differential reflection $\Delta R/R_0$, we measured the reflection spectrum $R_0$ from an excitation spot on our device away from the MoSe$_2$ layer. Together with our bare reflection spectrum $R$, we then calculated $\Delta R/R_0=\frac{R-R_0}{R_0}$. This  method of background subtraction is known to introduce systematic errors, due to the sensitivity of the reflection spectrum on e.g. the exact sample position with respect to the focal plane. Nevertheless we chose this approach, as the more accurate method of reconstructing the background from spectra at different gate voltages was not applicable due to the presence of resonances that energetically overlap at every measured doping level. 

\begin{figure}[h]
    \centering
    \includegraphics[width=\columnwidth]{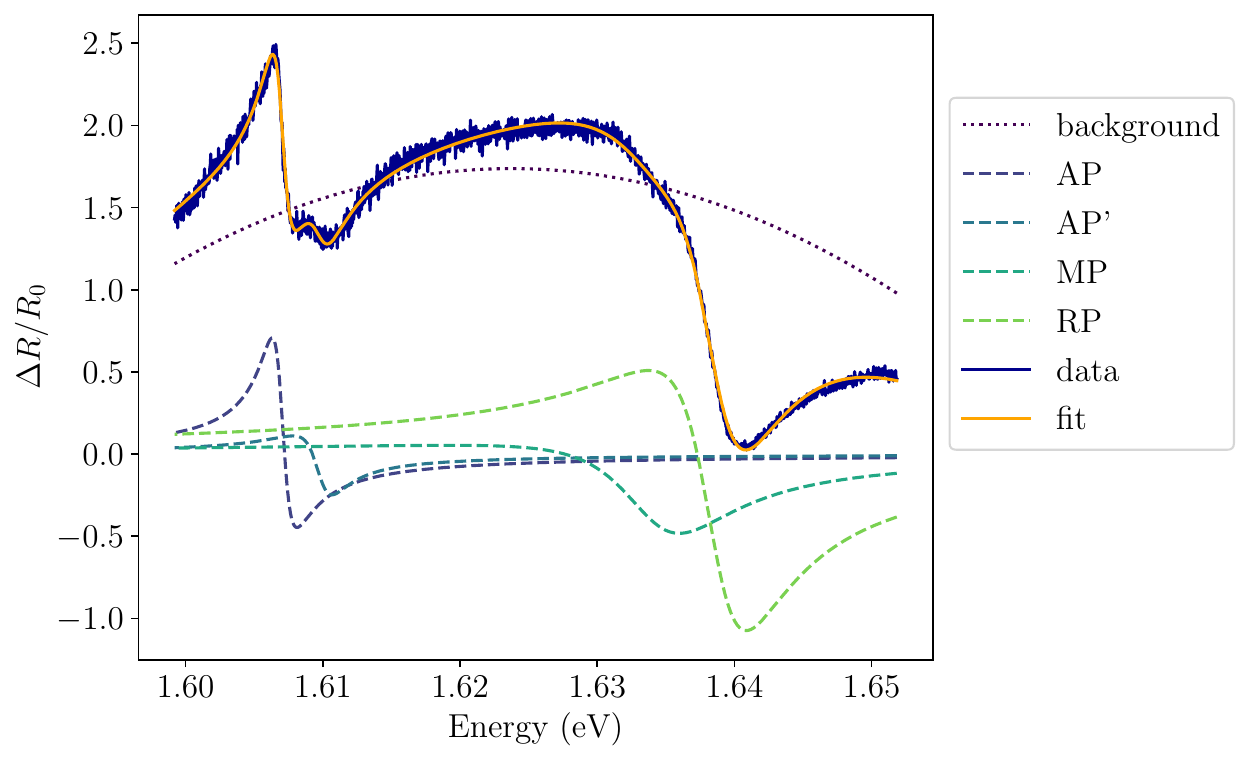}
    \caption{Example of fit to differential reflection spectrum at $V_\mu=1.56$~V ($\nu\approx1.25$). The data is plotted in dark blue, the dashed lines represent the components from the four different resonances. The dotted line shows the quadratic background. The orange solid line shows the total fit (sum of the four dispersive Lorentzians and background).}
    \label{fig:si_fitex}
\end{figure}

The shape of the differential reflection spectrum is not only determined by the reflection of the \mose\ layer, but also by its interference with the reflections from all other surfaces and interfaces within the stack and substrate. The reflection $R$ as a function of energy $E$ can be approximately described by the dispersive Lorentzian function
\begin{widetext}
\begin{equation}
    R(E,E_0,\gamma,A,\alpha)=A\cos(\alpha)\frac{\gamma/2}{(E_0-E)^2+(\frac{\gamma}{2})^2}+A\sin(\alpha)\frac{E_0-E}{(E_0-E)^2+(\frac{\gamma}{2})^2},
    \label{eq:displor}
\end{equation}
\end{widetext}
where $E_0$, $\gamma$, and $A$ are the resonance energy, linewidth, and amplitude, respectively. The parameter $\alpha$ denotes the wavelength-dependent phase shift that that arises from the interferences described above. 
In order to extract the oscillator strength $f=\pi A$, we fit the DR spectra with the total reflectivity
\begin{equation}
    R_{tot}(E)=\sum_pR(E,E_{0,p},\gamma_p,A_p,\alpha_p)+f_{bg}(E),
\end{equation}
where $p\in\{\text{X, \apone, \aptwo, \RP\, MP}\}$ denotes the resonance, and $f_{bg}(E)$ is a heuristic quadratic function necessary to remove a residual background due to imperfect background subtraction. We use the same background function for each spectrum in any gate-dependent DR dataset. Fig.~\ref{fig:si_fitex} shows an example of a DR spectrum at filling factor $\nu = 1.25$ and $B_\text{z}=0$~T, where all four resonances are present, with the fitted $R_{tot}$ and all its components.
Due to the difficulties regarding the proper treatment of the background subtraction described above, the quantitative value for the oscillator strength extracted from the fit functions is not meaningful, and only serves to qualitatively compare oscillator strengths of resonances taken within the same measurement run.

\begin{figure}[h]
    \centering
    \includegraphics[width=\columnwidth]{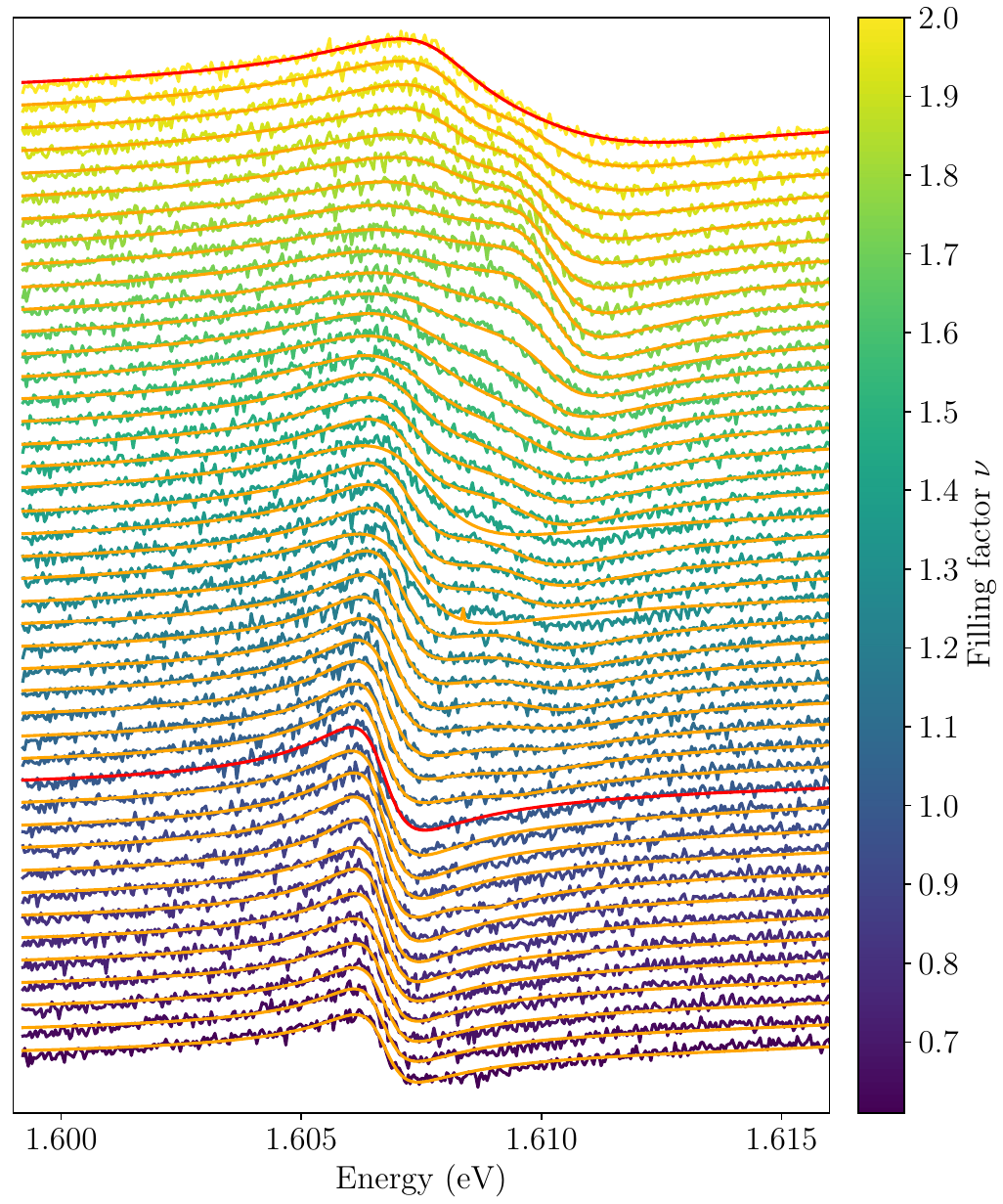}
    \caption{Differential reflection spectra showing the \apone\ and \aptwo\ resonances for filling factors between $\nu\approx0.6$ and $\nu=2$. The orange lines are fits to the spectra, red lines are fits at integer filling factors $\nu=1$ and $\nu=2$.}
    \label{fig:si_linecuts_ap12}
\end{figure}

Figure~\ref{fig:si_linecuts_ap12} shows DR spectra and fitted reflectivity for filling factors between $\nu\approx0.6$ and $\nu=2$, focusing on the \apone\ and \aptwo\ resonances.

\section{Charge Density and Moiré Length Calculation} \label{sec:app_ncalc}

\begin{figure}[h]
    \centering
    \includegraphics[width=\columnwidth]{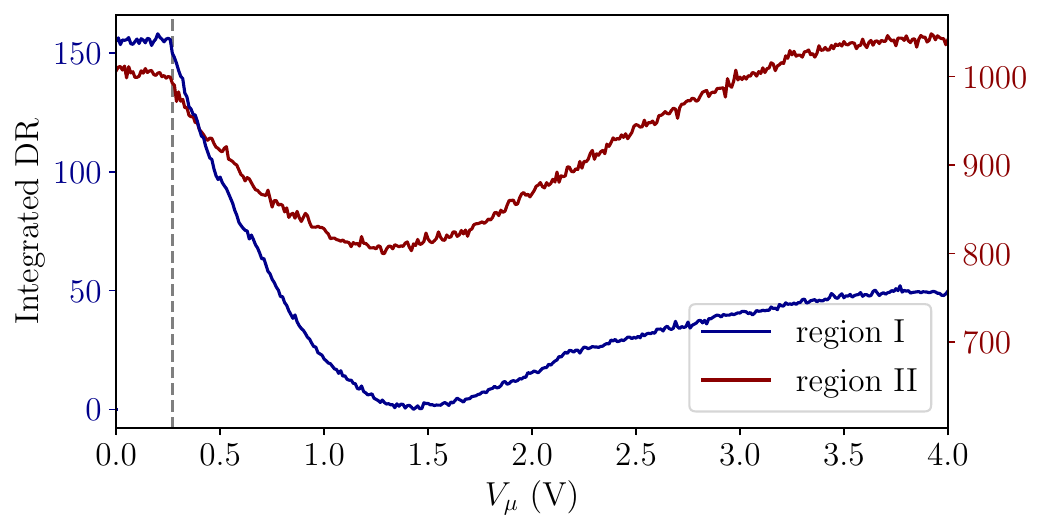}
    \caption{Integrated DR in the energy range {$1.626$~eV$\lesssim E\lesssim 1.652$~eV} for measurement spots in regions I and II.}
    \label{fig:si_integratedDR}
\end{figure}

We use a simple capacitor model to convert the voltages applied to top and bottom gates to a charge carrier density in the TMD layer. We model the device as two coupled capacitors, with the TMD as the central layer. The TMD layer is always grounded ($V_\text{TMD}=0$). The central layer is separated from the top (bottom) gate by a \hBN\ dielectric spacer with thickness $d_\text{top}=50\pm5$~nm ($d_\text{bot}=36\pm5$~nm). The thicknesses of the \hBN\ gate dielectrics are estimated from their colors in bright field microscopy. We apply the voltages $V_\text{TG}$ and $V_\text{BG}$ to the top and bottom gate, respectively. The charge density induced on the TMD layer through the capacitive coupling is thus
\begin{equation}
    \sigma = \varepsilon_\text{BN}\varepsilon_0\left(\frac{V_\text{TG}}{d_t}+\frac{V_\text{BG}}{d_b}-\left(\frac{1}{d_t}+\frac{1}{d_b}\right)V_0\right),
    \label{eq:sigma}
\end{equation}
where $\varepsilon_\text{BN}=3.4\pm0.2$ is the \hBN\ out-of-plane dc dielectric constant \cite{pierretDielectricPermittivityConductivity2022}, $\varepsilon_0$ is the vacuum permittivity, and $V_0$ is a constant that represents the finite voltage needed to surpass the charge gap of the TMD.  We determine $V_0$ from the integrated DR signal over the exciton / \RP\ resonance (see Fig.~\ref{fig:si_integratedDR}). As soon as charges are injected into the system, the exciton starts losing oscillator strength to the \AP, leading to a sharp cusp in the integrated DR. 

The charge carrier density is then $n=\sigma/e$ with $e$ the elementary charge.

As described in Sec.~\ref{sec:umklapp}, we determine the gate voltages at which unity filling of the moiré lattice ($V_\text{TG}|_{\nu=1}=V_\text{BG}|_{\nu=1}=1.49$~V) occur directly from the spectral features. We use Eq.~\ref{eq:sigma} to calculate the charge density at $\nu = 1$ and find $n|_{\nu=1}=1/A_\text{m}=(1.1\pm0.8)\cdot10^{12}$~cm$^{-2}$, where $A_\text{m}$ is the area of one moiré unit cell. The moiré length $a_\text{m}$ is then given by $a_\text{m}^2=\frac{2}{\sqrt{3}}A_\text{m}$. From this we can calculate the corresponding twist angle of the \tBN, which is given by $\theta_\text{m}=\arccos(1-\frac{a_\text{BN}^2}{2a_\text{m}^2})$, with $a_\text{BN}=2.5040$~\si{\angstrom} the lattice constant of \hBN\ \cite{lynchEffectHighPressure1966}.

\section{Photoluminescence spectroscopy} \label{sec:app_pl}

\begin{figure}[h]
    \centering
    \includegraphics[width = \columnwidth]{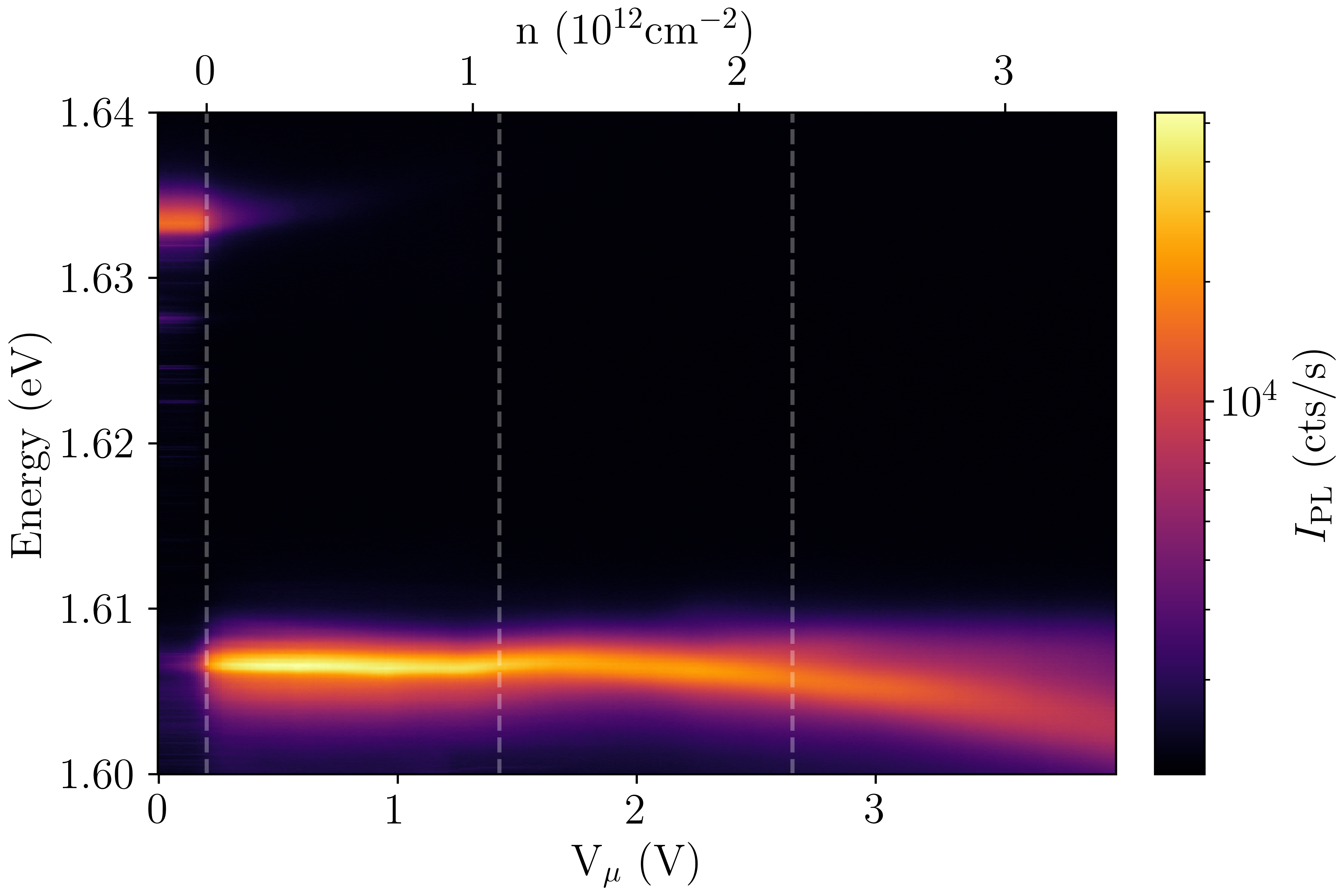}
    \caption{Photoluminescence spectra as a function of electron doping at $B_\text{z}=0$~T. The dashed gray vertical lines indicate the doping onset and integer fillings ($\nu=1$ and $\nu=2$) of the moiré lattice.}
    \label{fig:si_pl}
\end{figure}

We show doping-dependent photoluminescence (PL) spectra at $B_\text{z}=0$~T in Fig.~\ref{fig:si_pl}. We used a helium-neon laser at wavelength 632.8~nm (excitation power: 8.6~$\upmu$W before the insert) for the PL measurement. We observe a blue shift of the \AP\ at $\nu=1$, indicating an incompressible state of electrons.

\section{Umklapp resonances at $\nu=1/3$ and $\nu=2/3$} \label{sec:app_umklapp}

Since the Umklapp resonance can be understood as a high momentum exciton being scattered to zero momentum by a periodic repulsive potential, we can estimate the energy splitting $E_\text{U}$ between the Umklapp and the exciton / \RP\ by considering the band folding of the bare exciton dispersion at the edge of the periodic potential Brillouin zone. In our case, the repulsive periodic potential for the excitons stems from localized electrons in the moiré lattice. The bare exciton has both linearly and parabolically dispersing branches. As the linearly dispersing branch has a large slope, its Umklapp bands are far detuned in energy and therefore not relevant. The parabolic dispersion is given by $E(\mathbf{k})=\frac{\hbar^2\mathbf{k}^2}{2m_\text{X}}$, where $\mathbf{k}$ is the momentum and $m_\text{X}=1.2m_0$ is the exciton mass ($m_0$ is the free electron mass). The splitting is therefore given by $E_\text{U}=E(\textbf{k}_\text{U})$, where $k_\text{U}=|\textbf{k}_\text{U}|$ is the size of the Brillouin zone. For a triangular lattice with lattice constant $a$, we have $k_U=\frac{2\pi}{a}$. A triangular lattice with lattice constant $a$ corresponds to an electron density of $n=1/A$, where $A=\frac{\sqrt{3}}{2}a^2$ is the area of the real space unit cell. We therefore have $E_\text{U}(n)=\frac{4h}{m_\text{X}}n$.

\begin{figure*}[h]
\begin{minipage}[t]{0.75\textwidth}
    \vspace{0pt}
    \includegraphics[width=\textwidth]{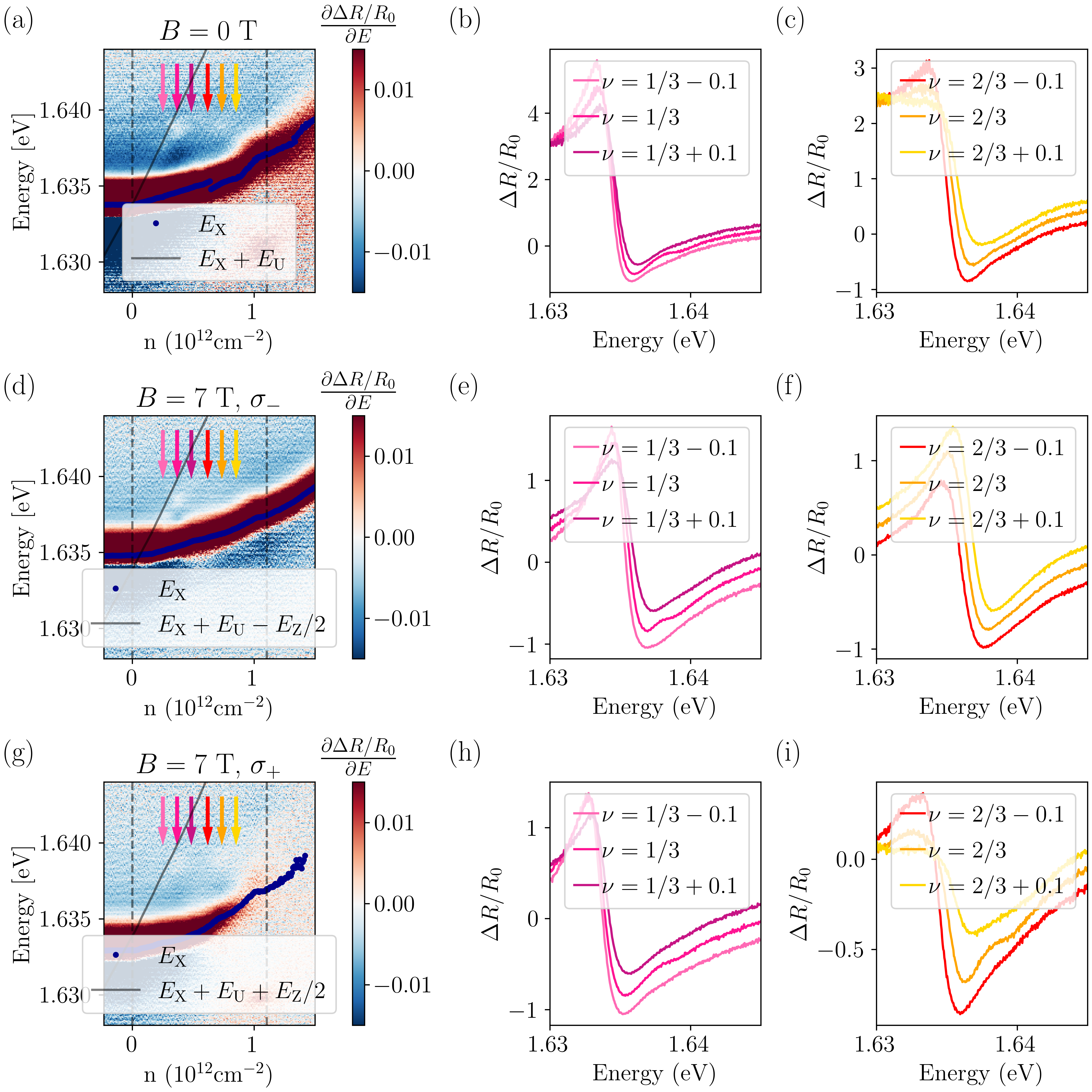}
  \end{minipage}\hfill
  \begin{minipage}[t]{0.25\textwidth}
    \vspace{0pt}
    \caption{
       Details of the Umklapp resonances at $B_\text{z}=0$~T (a-c), $B_\text{z}=7$~T in $\sigma_-$ polarization (d-f), and $B_\text{z}=7$~T in $\sigma_+$ polarization (g-i). Panels (a), (d), and (g) show differentiated DR around the exciton / RP resonance at low doping, as well as exciton / RP energy $E_\text{X}$ obtained from dispersive Lorentzian fits (blue dots), and Umklapp energy $E_\text{X}+E_\text{U}$ for triangular lattice as a function of $n$ (gray solid line). The gray dashed lines indicate the onset of doping and $\nu=1$. Panels (b), (e), and (h) show DR spectra at and around $\nu=1/3$, panels (c), (f), and (i) show DR spectra at and around $\nu=2/3$. An offset has been added to the spectra in panels (b), (c), (e), (f), (h), and (j) for clarity.
    } \label{fig:si_umklapp}
  \end{minipage}
\end{figure*}

At filling factors $\nu=1/3$ and $\nu=1$, we expect the electrons in the \tBN\ moiré lattice to form triangular Mott--Wigner states. In Fig.~\ref{fig:si_umklapp}(a,d,g), we show differentiated DR spectra, as well as the expected Umklapp energy $E_\text{X}+E_\text{U}$. For the data at $B_\text{z}=7$~T, we also take the Zeeman splitting $E_\text{Z}$ of the exciton into account. We see that the $\nu=1/3$ Umklapp resonance is at the expected energy.

At filling factor $\nu=2/3$, the situation is more complicated, as the electrons are expected to form a hexagonal lattice. The unit cell of this hexagonal lattice has the same size as the triangular lattice formed by the electrons at $\nu=1/3$, but has two occupied sites per unit cell. We therefore expect to see two Umklapp resonances centered around $E_\text{X}|_{\nu=2/3}+E_\text{U}|_{\nu=1/3}$. A small splitting between the two Umklapp resonances could arise due to hybridization between the two back-folded states, but we expect this splitting to be small compared to $E_\text{U}$. See also \cite{salvadorOpticalSignaturesPeriodic2022} for a detailed discussion of Umklapp resonances from triangular and hexagonal lattices. The Umklapp resonance we observe at $\nu=2/3$ indeed appears at roughly half the splitting expected for a triangular lattice, confirming that the $\nu=2/3$ Mott--Wigner state is hexagonal. We do not observe two Umklapp peaks at $\nu=2/3$, indicating either that the two Umklapp bands do not hybridize or the splitting is smaller than the Umklapp linewidth.

We note that we do not see the $\nu=2/3$ Umklapp in $\sigma_-$ polarization at $B_\text{z}=7$~T. We speculate that this is due to the closeness in energy to the much stronger, blueshifting exciton resonance.

\section{Hole side differential reflection at $B_\text{z}=0$~T and $B_\text{z}=7$~T} \label{sec:app_holeside}

\begin{figure}[h]
    \centering
    \includegraphics[width=\columnwidth]{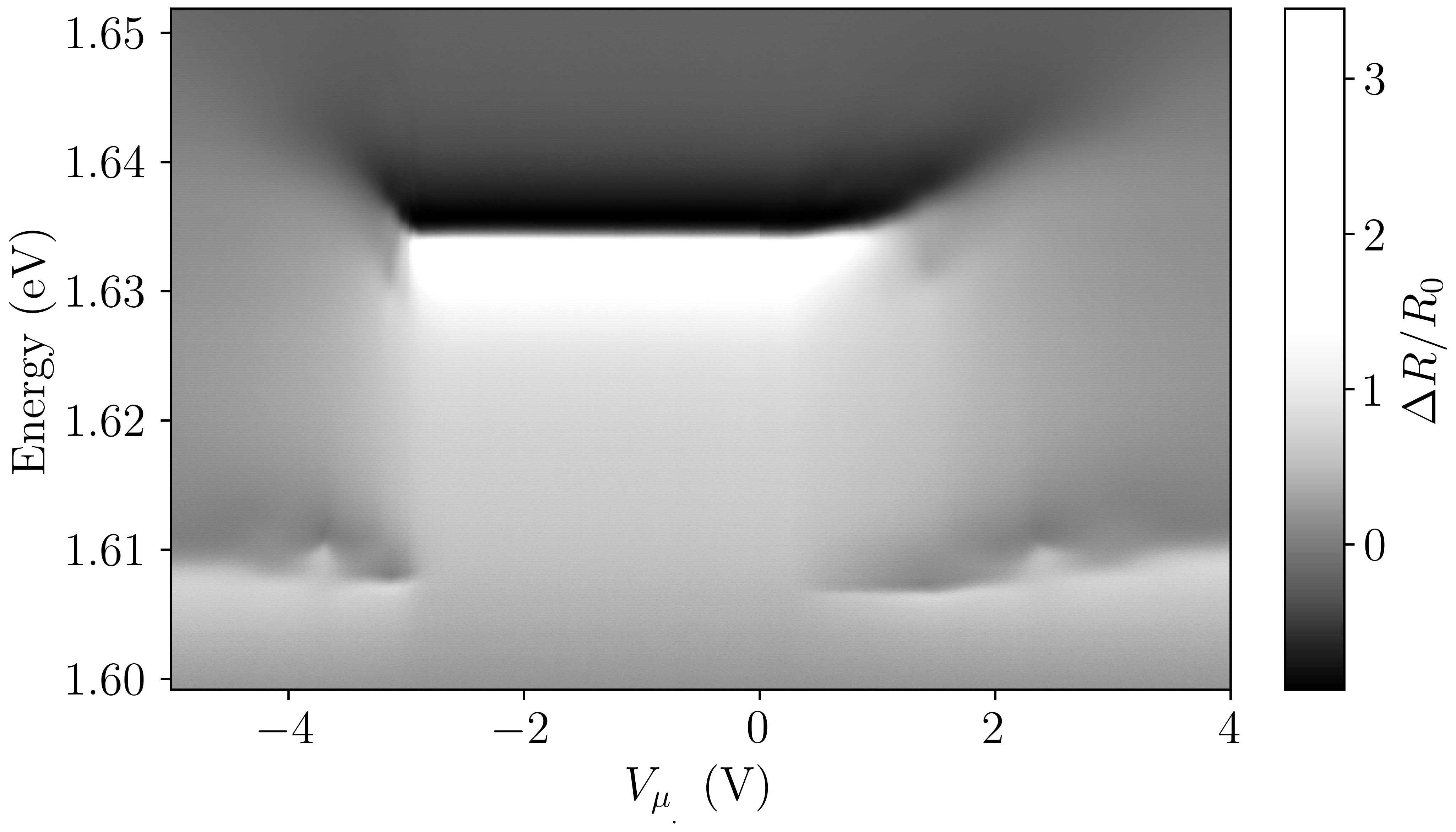}
    \caption{Differential reflection in region I as a function of $V_\mu$ (for $V_E=0$), showing qualitatively equal signatures of the moiré-trapped charge carriers for hole and electron doping.}
    \label{fig:si_holeside}
\end{figure}

As mentioned in the main text, we expect the moiré potential for electrons and holes generated by \tBN\ to have equal shapes and depths, the only difference being an offset in the plane by half a moiré unit cell (or equivalently, a change of sign). We indeed see the same spectral features that we attribute to the moiré potential (namely, the MP and \aptwo\ resonances) on the hole doping side as well, see Fig.~\ref{fig:si_holeside}. We concentrate on the electron side for our analysis, as for low doping, the charge carrier density dependence on the gate voltage is nonlinear due to poor contact quality for holes. This nonlinear hole doping behavior has been previously observed in moiré-free graphene-contacted \mose\ devices \cite{smolenskiInteractionInducedShubnikovHaas2019}, therefore we do not believe it to be a consequence of the moiré potential or proximity to \tBN\ layers.

\begin{figure}[h]
    \centering
    \includegraphics[width=\columnwidth]{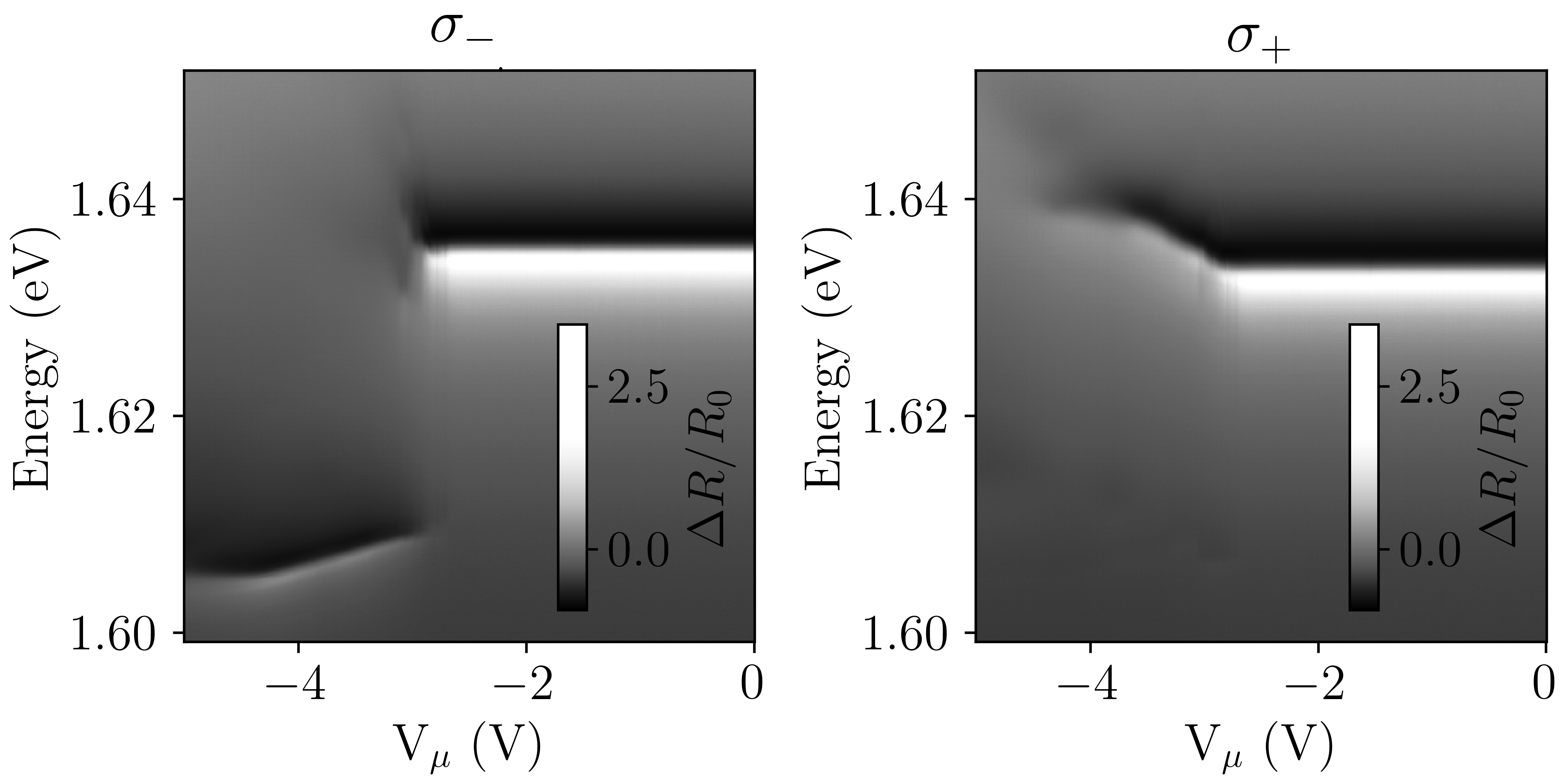}
    \caption{Differential reflection as a function of $V_\mu$ at $B_\text{z}=7$~T on the hole doping side.}
    \label{fig:si_hole_7T}
\end{figure}

We also show the DR spectra on the hole doping side at $B_\text{z}=7$~T in Fig.~\ref{fig:si_hole_7T}. Here, we again see the MP resonance and a single \AP\ in $\sigma_-$ polarization (probing the undoped valley), as well as a cusp in the $\sigma_+$-polarized exciton. In contrast to the electron side, the \AP\ exhibits a strong redshift.

\section{Effect of out-of-plane electric field} \label{sec:app_efield}

\begin{figure}[h]
    \centering
    \includegraphics[width=\columnwidth]{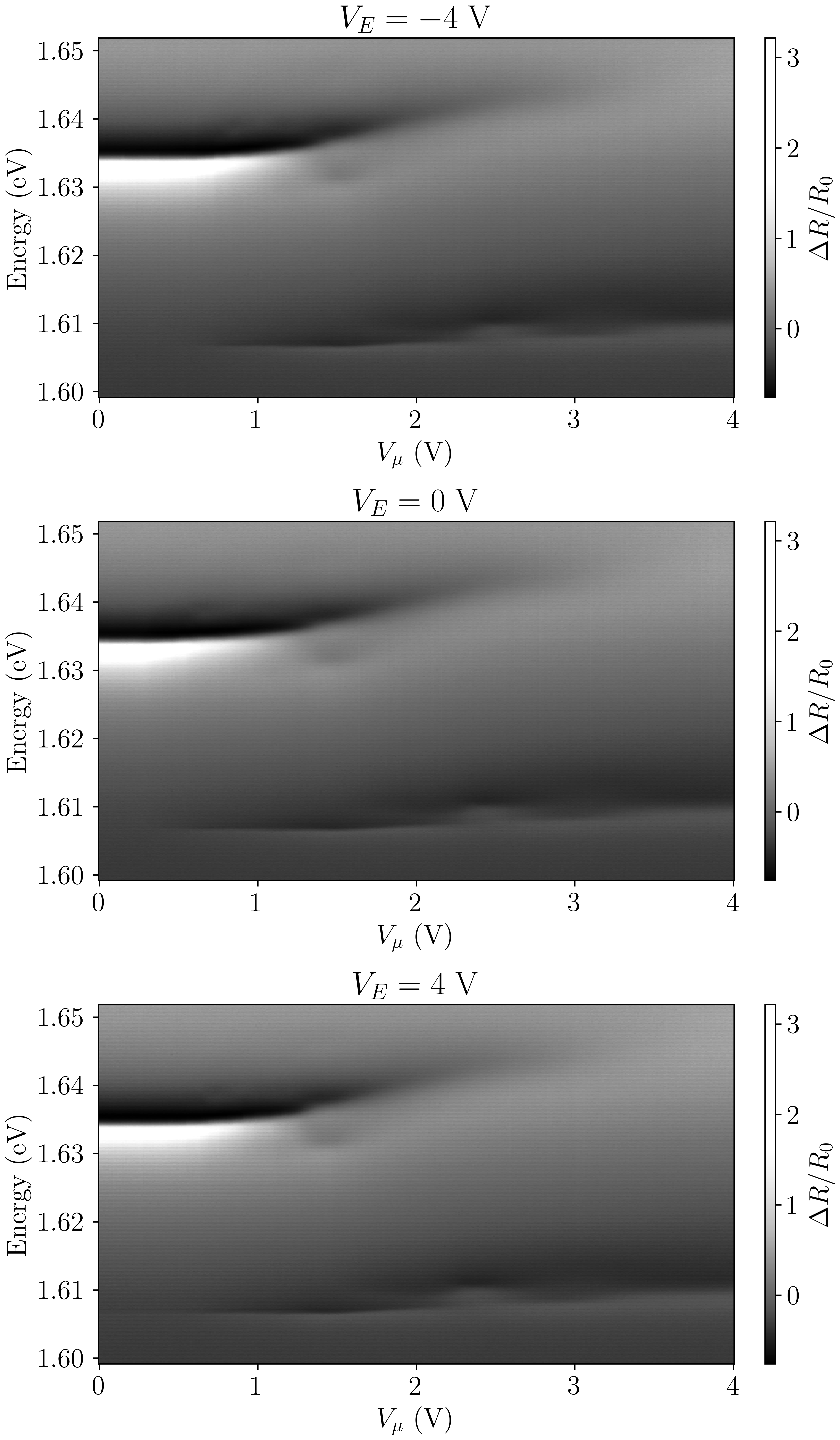}
    \caption{Differential reflection spectra in region I as a function of $V_\mu$ for three different values of $V_E$, showing the absence of an effect of applied out-of-plane electric field on the spectral signatures.  }
    \label{fig:si_vedep}
\end{figure}

Some previous studies report a hysteretic electric field dependence of ferroelectric \tBN\ domains due to field-induced switching of the polarization via an in-plane layer sliding transition \cite{viznersternInterfacialFerroelectricityVan2021, kimHarnessingMoireFerroelectricity2024}. We do not observe any qualitative effect of an applied out-of-plane electric field, as can be seen in Fig.~\ref{fig:si_vedep}, where we show DR spectra as a function of $V_\mu$ for $V_E=0.5(V_\text{TG}-V_\text{BG})=-4,0,4$~V, corresponding to fields $E=47,0,-47$~mV/nm, respectively. We assume that we are either unable to reach the critical electric field necessary to switch the ferroelectric state due to the breakdown limit of our gate dielectric, or that the relatively small-scale  moiré pattern stabilizes the domains.

\section{Theoretical model for trion binding in moiré} \label{sec:app_trionbinding}

\subsection{Model}
 
To model the trion binding energy in the presence of the twisted \hBN\ moiré we consider a rigid 1s exciton interacting with an electron that is subject to a periodic electrostatic potential. The Hamiltonian reads
\begin{align} \label{Eq:model}
    H = \frac{\pv^2_\text{X}}{2m_\text{X}} + \frac{\pv_\text{e}^2}{2m_\text{e}} + U( \rv_\text{e}) + V( \rv_\text{X} -\rv_\text{e}),
\end{align}
where $\pv_\text{X}$ and $\pv_\text{e}$ denote the momentum operators for the exciton (X) and electron (e), respectively. The exciton and electron masses are represented by $m_\text{X}$ and $m_\text{e}$, and are set to $1.2m_0$ and $0.7m_0$ with $m_0$ being the bare electron mass. The terms $U(\rv_\text{e} )$ and $V( \rv_\text{X} -\rv_\text{e})$ denote the moiré potential and the exciton-electron interaction potential, respectively.

Inspired by past treatments of moiré potentials in \TMD s~\cite{wuTopologicalExcitonBands2017}, we approximate the moiré potential through its lowest harmonic expansion 
\begin{align} \label{Eq:moirePot}
    U( \rv_\text{e}) =  \sum_{j=1}^6 U_j e^{i \bv_j\cdot \rv_\text{e}  },
\end{align}
with $U_j$ as the harmonic coefficients and $\bv_j$ the reciprocal lattice vectors of the first shell. These vectors can be derived by applying a $(j-1)\times\pi/3$ rotation to $\bv_1 = (\frac{4\pi}{\sqrt{3}a_\text{m}}, 0)$, where $a_\text{m}$ is the experimentally determined moiré length of $10.2$~nm.

To ensure that the moiré potential is real and exhibits three-fold rotational symmetry, the harmonic coefficients must satisfy:
\begin{align}
    U_1 = U_3 = U_5, \quad U_2 = U_4 = U_6, \quad U_1 = U_4^*.
\end{align}
This constraint narrows the form of the expansion coefficients to $U_1 = U_0 e^{i \psi}$, with $\psi = \pi/2$ producing a moiré potential that aligns with the expected form of twisted \hBN\ near a $0^\circ$ twist angle. The modulation of the moiré potential $\Delta_\text{m}$ discussed in the main text is quantified by $\Delta_\text{m} = 6 \sqrt{3} U_0$.

For the exciton-electron interaction, we adopt the effective interaction~\cite{amelioPolaronSpectroscopyBilayer2023}
\begin{align} \label{Eq:ExEInteraction}
    V(\rv) = - \frac{W_0}{(a_{\rm X}^2 + |\rv|^2)^2}
\end{align}
where $a_{\rm X}$ is the 1s exciton radius of approximately 1.1 nm~\cite{gorycaRevealingExcitonMasses2019}. We have checked that the choice of exciton-electron interaction (e.g., Gaussian with finite range, or contact) does not affect the results presented in the main text: the minor quantitative differences are well within the uncertainty in the trion binding in the absence of moiré.

\subsection{Trion binding in moiré}
To determine the trion binding energy in the presence of the moiré potential, we take advantage of the relatively weak coupling between the relative and center-of-mass (CoM) coordinates; the coordinates exactly decouple in the absence of moiré. Consequently, we reformulate the Hamiltonian presented in Eq.~\eqref{Eq:model} as follows:
\begin{align}
    H = H_{\rm rel} + H_{\rm CoM} + H_{\rm rc},
\end{align}
where the system is decomposed into:
\begin{subequations}
    \begin{align}
    H_{\rm rel} &= \frac{\pv^2}{2 m_r} + V(\rv)\\
    H_{\rm CoM} &= \frac{\vb{P}^2}{2M} + U(\Rv) \\
    H_{\rm rc} &= U\left(  \Rv - \frac{m_\text{X}}{M} \rv \right) - U(\Rv). \label{Eq:Hrc}
\end{align}
\end{subequations}
In these expressions, $\Rv$ and $\rv$ denote the CoM and relative coordinates, respectively, with $\vb{P}$ and $\pv$ representing their corresponding momentum. The masses $M$ and $m_r$ are defined as the total mass and the reduced mass of the exciton-electron system, respectively. The relationship between the coordinate system is summarised by:
\begin{subequations}
    \begin{align}
    \Rv &= \frac{m_\text{X} \rv_\text{X} + m_\text{e} \rv_\text{e}}{M}, \\
    \rv &= \rv_\text{X} - \rv_\text{e}, \\
    \pv &= \frac{m_\text{e} \pv_\text{X} - m_\text{X} \pv_\text{e}}{M}, \\
    \vb{P} &= \pv_\text{X} + \pv_\text{e}.
\end{align}
\end{subequations}
This decomposition demonstrates that if the expectation value of $\rv$ is significantly smaller than the moiré length $a_\text{m}$, then the coupling term $H_{\text{rc}}$ in Eq.~\eqref{Eq:Hrc} becomes negligible. This motivates us to construct a basis using the eigenstates of the relative and CoM motions.

\subsubsection{Relative motion}
In polar coordinates, the Hamiltonian of the relative motion reads 
\begin{align}
    H_{\rm rel} = - \frac{\hbar^2}{2m_r} \left( \pdv[2]{}{r} + \frac{1}{r} \pdv{}{r} - \frac{1}{r^2} \pdv[2]{}{\theta} \right) + V(r).
\end{align}
Given the spherical symmetry of the interaction potential, angular momentum is conserved by $H_{\rm rel}$. The eigenstates therefore take the form $\psi(r,\theta) e^{-i l \theta}/\sqrt{2 \pi}$, where $l=0, \pm1 , \dots$, facilitating the discretization and efficient solution to yield the eigenstates and eigenenergies, denoted by $\ket{\psi_{l n}}$ and $E^r_{ln}$, respectively. We point out that the inclusion of the moiré potential breaks the rotational symmetry, and therefore we must include multiple angular momentum channels in our basis for the trion wave function.

\subsubsection{Centre-of-mass motion}
The CoM Hamiltonian can be efficiently solved using Bloch's theorem. In particular, at zero CoM quasi-momentum, we express the eigenvalue problem as
\begin{align}
    H_{\rm CoM} \ket{\nu} = E_\nu \ket{\nu},
\end{align}
where
\begin{align}
    \ket{\nu} = \sum_{\bv} \phi^{(\nu)}_{\bv} \ket{\bv},
\end{align}
with the sum over all reciprocal lattice vectors. We point out that CoM quasi-momentum is a conserved quantity and we henceforth always set it to zero. We can index the reciprocal lattice vectors uniquely by the vector $\mv = (m,n)$ through $\bv_{\mv} = m \bv_1 + n \bv_6$. Then the expansion coefficients satisfy the eigenvalue problem
\begin{widetext}
	\begin{align}
		E \phi^{(\nu)}_{\mv}  &= \frac{| \bv_{\mv}|^2}{2 M }  \phi^{(\nu)}_{\mv} + U_1  \phi^{(\nu)}_{(m-1,n)} + U_3 \phi^{(\nu)}_{(m,n+1)} \nonumber  + U_5 \phi^{(\nu)}_{(m+1,n-1)} \nonumber \\
		&{} \qquad +  U_2  \phi^{(\nu)}_{(m-1,n+1)} +  U_4 \phi^{(\nu)}_{(m+1,n)} +  U_6 \phi^{(\nu)}_{(m,n-1)},
	\end{align}
\end{widetext}
where $\phi^{(\nu)}_{\mv} \equiv \phi^{(\nu)}_{\bv_{\mv}}$.

 \subsubsection{Eigenvalue problem for trion binding}
The final ingredient required for our analysis is the matrix elements of relative-CoM coupling, which are given by
\begin{widetext}
    \begin{align}
        \bra{ \psi_{nl}, \nu} H_{\rm rc} \ket{\psi_{n' l'}, \nu'} =  \sum_{j=1}^6 U_j \sum_{\bv}  \int dr^2 \psi_{nl}^{*}(\rv) \psi_{n'l'}(\rv)  \phi^{\nu}_{\bv}  \phi^{\nu'}_{\bv - \bv_j} \left(e^{-m_\text{X} \bv_j \cdot \rv/M}-1 \right).
    \end{align}
\end{widetext}
To solve for the trion's ground state, we expand the complete Hamiltonian in terms of a general two-body wave function with zero CoM quasi-momentum:
\begin{align}
    \ket{\Psi} = \sum_{nl\nu} \zeta^\nu_{nl}   \ket{ \psi_{nl}, \nu}.
\end{align}
Here, the expansion coefficients satisfy
\begin{align} \label{Eq:EigProbComplete}
    E \zeta^\nu_{nl} = (E^r_{nl} + E_\nu)\zeta^\nu_{nl} + \sum_{n'l' \nu'} \bra{ \psi_{nl}, \nu} H_{\rm rc} \ket{\psi_{n' l'}, \nu'}  \zeta^{\nu'}_{n'l'}.
\end{align}
Given that the matrix elements of $H_{\rm rc}$ are small, the ground state converges quickly in this basis. Equation~\eqref{Eq:EigProbComplete} allows us to determine trion binding in the presence of moiré, and to explore the effect of a finite range interaction on this binding, as discussed in the main text.

\begin{figure}
    \centering
    \includegraphics[width=\columnwidth]{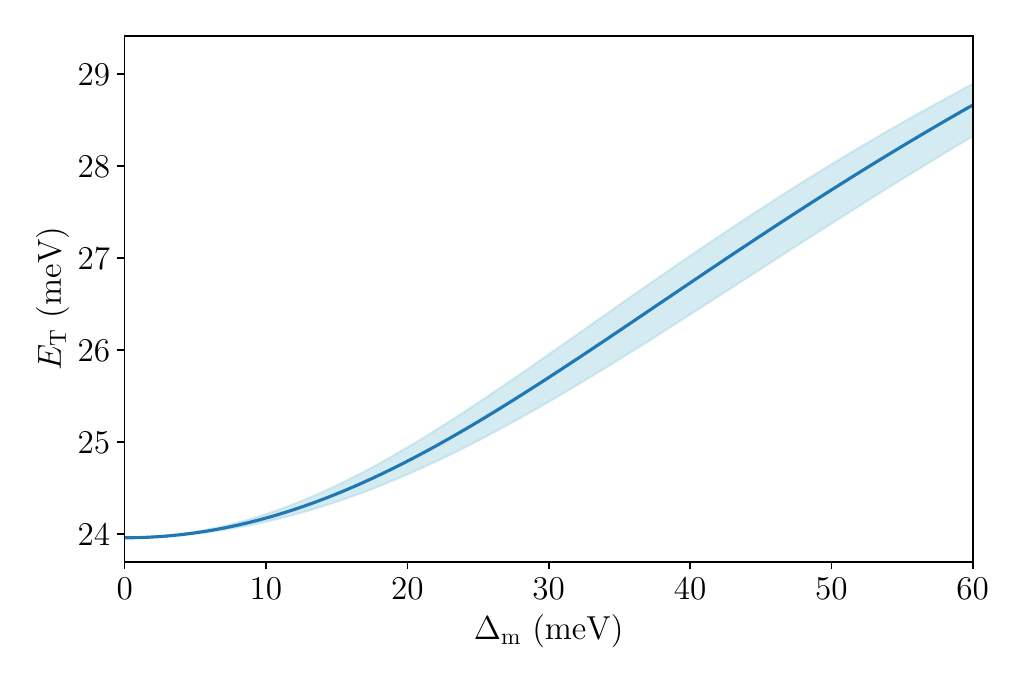}
    \caption{Trion binding energy as a function of moiré potential depth, where the trion in the absence of moiré has binding 23.96 meV. The blue shading shows the uncertainty in the trion binding introduced by the uncertainity in the moiré length ($a_{\rm m} = 10.2 \pm 0.7$ nm).}
    \label{fig:si_theory}
\end{figure}

Figure~\ref{fig:si_theory} shows the trion binding as a function of moiré depth. Here we also show the uncertainty in the moiré length by showing the range of trion binding energies (i.e., the blue shading). Importantly, the uncertainty from the uncertainty in the moiré length is not as large as the uncertainty in the uncertainty in the trion binding in the absence of moiré.

\section{Theoretical model for electrons and excitons at \texorpdfstring{$\bm{\nu=2}$}{TEXT}} \label{sec:app_theorynu2}

\begin{figure}[h]
    \centering
    \includegraphics[width=\linewidth]{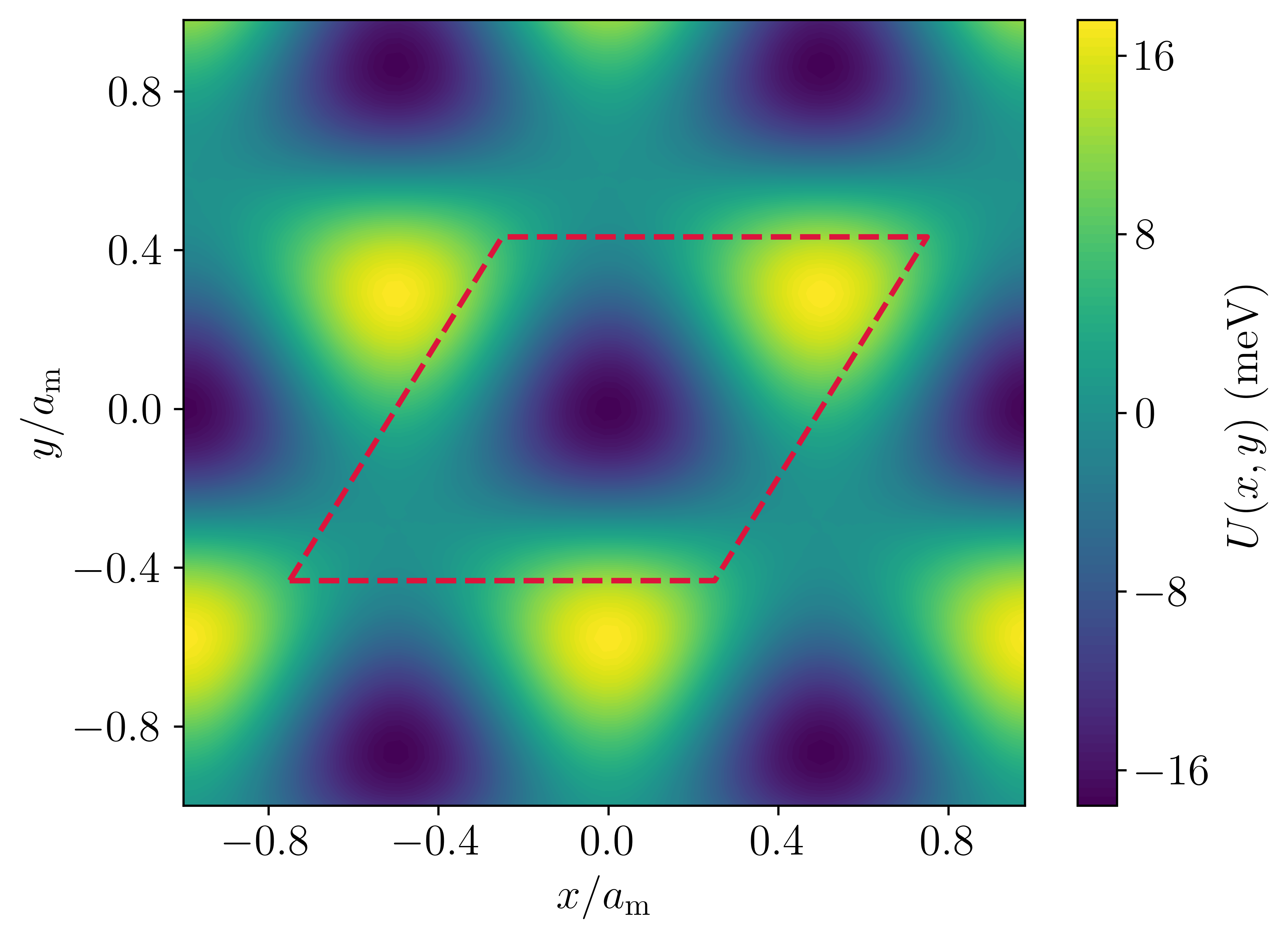}
    \caption{The moiré potential with a depth of 35 meV, which is chosen to yield the experimentally determined trion binding energy. The red parallelogram shows the moiré unit cell; we enforce homogeneous Dirichlet boundary conditions for all particles along this boundary in the exact diagonalization calculations.}
    \label{fig:moirePot}
\end{figure}

We now provide details on our theoretical model for the study of electrons, and their binding to an exciton, at $\nu=2$. With two electrons per site, we expect the formation of a band insulator, albeit with strongly modified bands from strong electron correlations. Inspired by this, our model completely neglects hopping between sites, and hence we treat the electrons as localized to individual moiré unit cells; neighboring sites are accounted for via a Hartree potential. Since the electrons are confined to unit cells, we do not account for exchange interactions. Figure~\ref{fig:moirePot} shows the theoretical moiré potential and the corresponding unit cell, which by construction is centered around a minimum of the potential. We begin our discussion by considering the singlet and triplet energy separation at $\nu=2$, and then proceed to the three body problem, for two electrons and an exciton in a single unit cell.

\subsection{Singlet-Triplet energy difference}
At $\nu=2$ we have two electrons per moiré unit cell with Hamiltonian
\begin{widetext}
    \begin{align} \label{Eq:hamnu2}
    H^{\rm{e}}_{\nu=2} = \frac{\pv_{\rm{e}1}^2}{2 \me} + U(\rveone) + \VH(\rveone) + \frac{\pv_{\rm{e}2}^2}{2\me} + U(\rvetwo) + \VH(\rvetwo) + V_{\rm RK}(\rveone - \rvetwo)
    \end{align}
\end{widetext}

where both electrons (1 and 2) enter the Hamiltonian identically. Here $U(\rve)$ is defined in Eq.~\eqref{Eq:moirePot} and $ \VH(\rve)$ is the Hartree potential from  electrons on neighboring sites, which is given by
\begin{align} \label{Eq:HartreePot}
    \VH(\rve) = \int_{\rv' \notin \rm \UM} d^2r' \, V_{\rm RK}(\rve - \rv') n_\text{e} (\rv').
\end{align}
Here we have introduced the electronic density $n_\text{e}(\rv)$, the moiré unit cell $\UM$ and the Rytova-Keldysh potential
\begin{align}
    V_{\rm RK}(\rv) = \frac{e^2}{8 \varepsilon_0 r_0} \left[ H_0 \left( \frac{\kappa |\rv|}{r_0} \right) - Y_0 \left(\frac{\kappa |\rv|}{r_0}  \right) \right],
\end{align}
where $H_0$ and $Y_0$ are the Struve functions and Bessel functions of the second kind, respectively; $\kappa \approx 4.4$ is the average dielectric constant of the encapsulating materials; $r_0 \approx 3.4$ nm is the screening length; $e$ is the electric charge; and $\varepsilon_0$ is the vacuum permittivity~\cite{gorycaRevealingExcitonMasses2019}.

Since we assume that the electron wavefunctions in each moir\'e unit cell are identical (which defines $n_\text{e}$), we diagonalize the Hamiltonian in Eq.~\eqref{Eq:hamnu2} iteratively to reach self-consistency. Note that in performing the diagonalization we must take homogeneous Dirichlet boundary conditions such that the electrons are entirely localized in a moiré unit cell. It should also be noted that as presently shown the integral in Eq.~\eqref{Eq:HartreePot} is not convergent due to the long-range nature of the Rytova-Keldysh potential (which behaves asymptotically like the Coulomb potential). We regularize this divergence by replacing $V_{\rm RK}(\rv) \to V_{\rm RK}(\rv) e^{- |\rv| / \tilde r}$, which removes the long range tail of the potential. While the choice of $\tilde r$ will always shift the absolute energies (which is not relevant to our results), for sufficiently large $\tilde r$  the wave functions and relative eigenenergies of the Hamiltonian in Eq.~\eqref{Eq:hamnu2} are not affected.

\begin{figure}
    \centering
    \includegraphics[width = \linewidth]{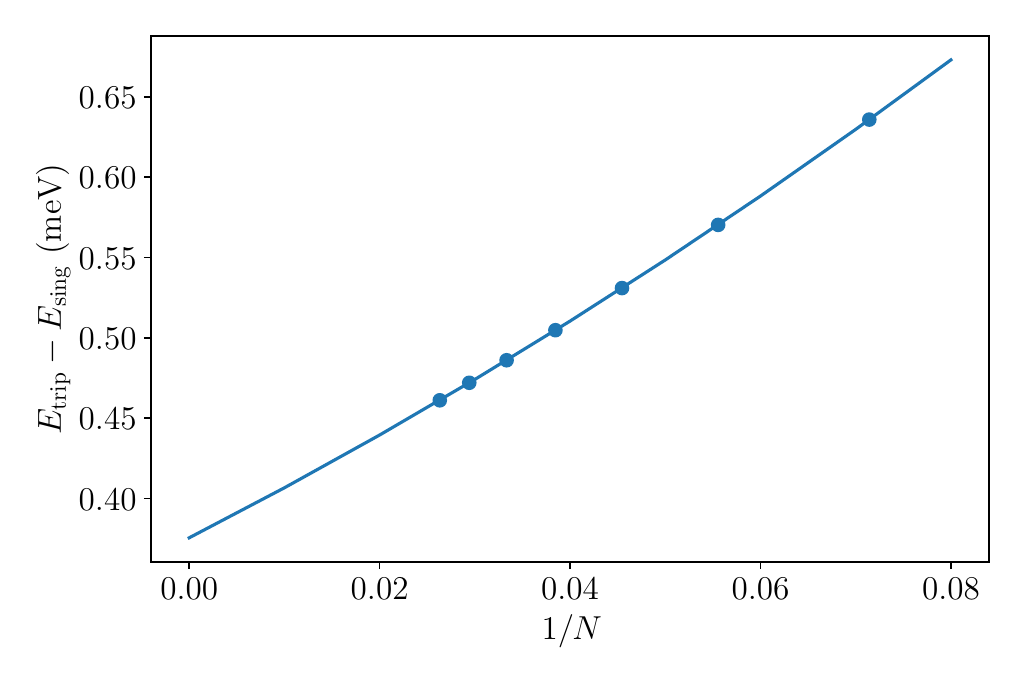}
    \caption{Triplet to singlet energy difference as a function of the inverse number of grid points along each dimension ($N$). The triplet is always higher in energy, but the difference to the singlet decreases with increasing $N$. The numerical data is well fit by a quadratic polynomial, which allows us to extrapolate to $1/N=0$, where the triplet-singlet energy difference is 0.38 meV.}
    \label{fig:TripSingEng}
\end{figure}

The eigenstates of Eq.~\eqref{Eq:hamnu2} have a definite parity under exchange: states that are even (odd) under exchange are associated with the spin singlet (spin triplet).
By discretizing and iteratively solving Eq.~\eqref{Eq:hamnu2}, we find that the singlet is the ground state, but the triplet energy is only slightly higher.
Here, we emphasize that when solving for either the singlet or triplet wave function, we assume that all other sites are occupied with the same singlet or triplet wave function, respectively.
Figure~\ref{fig:TripSingEng} shows the singlet-triplet energy difference as a function of the inverse number of grid points along each dimension of the unit cell $N$. The numerical data is well fit by a quadratic polynomial (i.e. $c_0 + c_1/N + c_2 /N^2$ with $c_i$ the fitting parameters), which motivates us to extrapolate to $1/N=0$, where the energy difference is 0.38 meV.

\begin{figure}
    \centering
    \includegraphics[width=\linewidth]{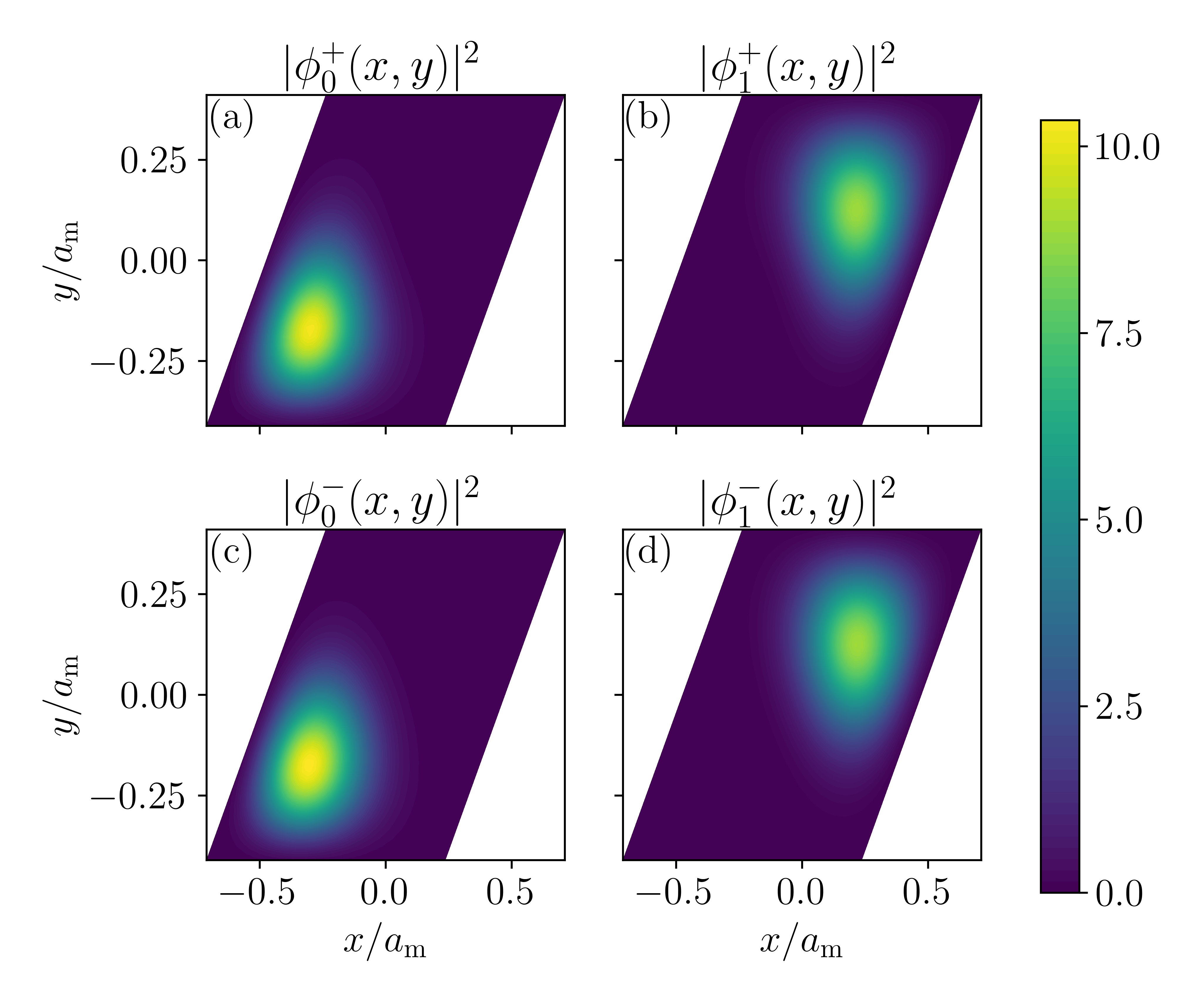}
    \caption{Orbitals in the decomposition of the singlet (top row) and triplet (bottom row) for two electrons confined to a single moiré unit cell. The orbitals of the singlet and triplet have striking similarity owing to the fact that the Coulomb interaction dominates over the moiré energy.}
    \label{fig:TripSingOrbitals}
\end{figure}

In order to understand this small energy difference, we decompose the singlet ($\Psi^+$) and triplet $(\Psi^-)$ wave functions into single-particle orbitals. In particular, we find decompositions of form
\begin{subequations}
    \begin{align}
    \Psi^+(\rveone,\rvetwo) &\simeq \frac{1}{\sqrt{2}}\left(\phi^{+}_0(\rveone)  \phi^{+}_1(\rvetwo)+\phi^{+}_1(\rveone)  \phi^{+}_0(\rvetwo) \right)\\
    \Psi^-(\rveone,\rvetwo) &\simeq \frac{1}{\sqrt{2}}\left(\phi_0^{-}(\rveone)\phi_1^{-}(\rvetwo)-\phi_1^{-}(\rveone)\phi_0^{-}(\rvetwo) \right).
\end{align}
\end{subequations}
In both cases, the overlap of the approximation decomposition and the exact eigenstate exceeds 0.98. Figure~\ref{fig:TripSingOrbitals} shows the orbitals in the singlet and triplet cases, which are seen to be remarkably similar. This is a consequence of the Coulomb interaction dominating over other energy scales, such that the two electrons seek to avoid each other in the moiré unit cell. Consequently, even in the singlet case, the electrons seek out near-orthogonal orbitals, similar to those of the triplet.

The small singlet-triplet energy separation suggests that at the temperatures of the experiment ($k_B T_{\rm exp} \approx 0.36$ meV) the electrons form a mixture of singlets and triplets across the various moiré unit cells. Owing to this, in what follows we modify our iterative procedure by always defining the Hartree potential with the assumption that the neighboring sites host singlets. Assuming instead that the neighboring sites host triplets, or a mixture, makes no qualitative difference to the findings and only minor quantitative differences.

\subsection{Exciton-electron bound states}

We now explore bound states between an exciton and two electrons. To this end, we first introduce the Hamiltonian for a free exciton
\begin{align}
    H_{\rm X} = \frac{\pv_{\rm X}^2}{2\mX},
\end{align}
where we once again enforce homogeneous Dirichlet boundary conditions, and we denote the ground state wave function and energy by $\psi_{\rm X}^{(0)}(\rv_{\rm X})$ and $E^{(0)}_{\rm X}$, respectively.
The three-body Hamiltonian is then given by
\begin{align} \label{Eq:ThreeBodyHam}
    H^{\rm eX}_{\nu=2} = H^e_{\nu=2} +H_{\rm X} + V_{\sigma_1}(\rveone - \rvX) + V_{\sigma_2}(\rvetwo - \rvX).
\end{align} 
Here, we take the exciton-electron interaction to be spin-dependent, thus phenomenologically accounting for how the spin of the electron inside the exciton affects its interaction with $\uparrow$ versus $\downarrow$ electrons. For simplicity, but without loss of generality, we take the electron inside the exciton to be spin $\downarrow$. Then, the exciton-electron interaction is given by
\begin{subequations}
    \begin{align}
        V_\uparrow (r) &= -\frac{W_0}{(a_{\rm X}^2 + r^2)^2}\\
        V_\downarrow (r) &= -\frac{W_0}{(a_{\rm X}^2 + r^2)^2} + P_0 \exp(-r/a_{\rm X}).
    \end{align}
\end{subequations}
For $V_\uparrow$ we use the same potential that was employed in Eq.~\eqref{Eq:ExEInteraction}, where $W_0$ is again chosen to give the correct trion binding in the absence of moiré, but now with the homogeneous Dirichlet boundary conditions. Meanwhile, for $V_\downarrow$, we include a short-range repulsion, which imitates the Pauli exclusion principle. In particular, $P_0$ is chosen such that in the absence of moiré the trion is not bound. We find that the results are mostly insensitive to the precise value of $P_0$ as long as the potential supports no bound states.

Since the model in Eq.~\eqref{Eq:ThreeBodyHam} conserves all individual spins, we diagonalize the Hamiltonian in the individual spin basis $\{\ket{\uparrow \uparrow}, \ket{\uparrow \downarrow}, \ket{\downarrow \uparrow} \ket{\downarrow \downarrow} \}$. Resultingly, we find eigenstates and eigen-energies, which we denote by $\psi^{(\sigma \sigma)}_i(\rvX, \rveone, \rvetwo)$ and $E_i^{\sigma \sigma}$, respectively. 

We must be careful to ensure the eigenstates have the correct anti-symmetrization under electron exchange. In the case of $\ket{\uparrow \uparrow}$ we only include eigenstates that satisfy $\psi^{(\uparrow \uparrow)}_i(\rvX, \rveone, \rvetwo) = - \psi^{(\uparrow \uparrow)}_i(\rvX, \rvetwo, \rveone)$. Meanwhile, for differing spins the exchange operator no longer decomposes into a spin and orbital component; we instead construct anti-symmetrized eigenstates according to
\begin{align}
    \ket{\bar{\psi}^{(\uparrow \downarrow)}_i} = \frac{1}{\sqrt{2}} \left( \ket{\psi^{(\uparrow \downarrow)}_i} \ket{\uparrow \downarrow} -  \ket{\psi^{(\downarrow \uparrow)}_i} \ket{\downarrow \uparrow}  \right),
\end{align}
where we have used the fact that $\psi^{(\uparrow \downarrow)}_i(\rvX, \rveone, \rvetwo) = \psi^{(\downarrow \uparrow)}_i(\rvX,  \rvetwo,\rveone)$ and $E^{\uparrow \downarrow}_i = E^{\downarrow \uparrow}_i$.

In order to determine the oscillator strength of each bound state, we consider its overlap with the initial state: two electrons either in a singlet or triplet, and a zero-momentum exciton. In the initial state, in absence of magnetic field, the three triplet states are degenerate and symmetrically equivalent. Hence, we are free to choose our quantization axis at will, and for simplicity we choose the same basis as in our final state, where the direction is set by the electron spin in the exciton. An initial state with electrons forming the triplet $\ket{T_{+}}= \ket{\uparrow \uparrow}$ will only have overlap with states in the set $\{\ket{\bar \psi^{(\uparrow \uparrow)}_i} \}$ (where the bar denotes anti-symmetrized states), whereas an initial state with electrons in a singlet $\ket{S} = \frac{1}{\sqrt{2}} \left(\ket{\uparrow \downarrow} - \ket{\downarrow \uparrow}  \right)$ or the triplet $\ket{T_0} = \frac{1}{\sqrt{2}} \left(\ket{\uparrow \downarrow} + \ket{\downarrow \uparrow}  \right)$ will have overlap with states in the set $\{ \ket{\bar{\psi}^{(\uparrow \downarrow)}_i} \}$. We emphasize that while the exciton cannot flip the electron spin, it will still mix singlet and triplet states because the exciton becomes entangled with the electron spins.

Since the eigenstates associated with $\ket{\downarrow \downarrow}$ are not bound, they are irrelevant for the attractive polaron lines and we henceforth exclude them, and the associated initial state $\ket{T_-} = \ket{\downarrow \downarrow}$ from our analysis.

\begin{figure}
    \centering
    \includegraphics[width = \linewidth]{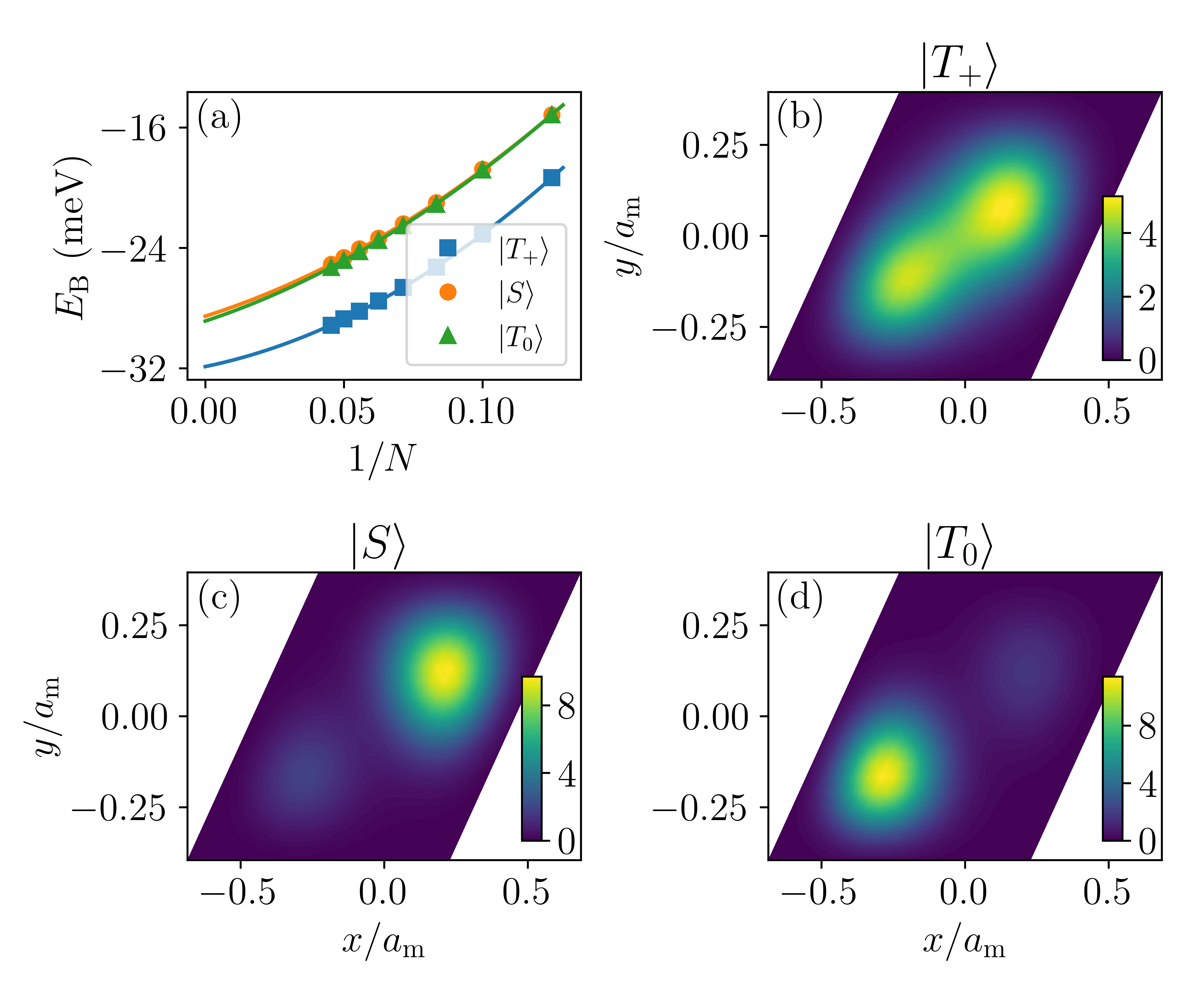}
    \caption{Comparison of the energies and wavefunctions of two electrons and an exciton confined to a moiré unit cell, where each bound state is identified by the spin-configuration of the associated initial state. \textbf{(a)} The binding energies as a function of $1/N$, which are well-fit by a quadratic polynomial, which allows us to extrapolate to $1/N=0$. \textbf{(b)--(d)} The wavefunction of the exciton for each of the three bound states (where the electrons have been traced out). In the case of $T_+$, where the exciton can attractively interact with both electrons (without phase space filling) one observes that the exciton delocalizes across both electrons.}
    \label{fig:threeBody}
\end{figure}

In general, we find that each initial state primarily has overlap with only one bound state within $\sim10$ meV of the ground state. In the case of $\ket{T_{+}}$ there is precisely one bound state with finite overlap (with magnitude squared 0.57). Meanwhile, in the case of the singlet or $\ket{T_{0}}$ there are two contributing bound states, but one dominates for each, which we show in Table~\ref{Tab:Overlaps}. This leads us to identify each bound state by its associated initial state (which we denote by the spin-configuration, i.e., $T_+$, $T_0$ or $S$). Furthermore, we measure the energy of each bound state with respect to its associated initial state (where we take the energy of the exciton in the initial state as $E^{(0)}_{\rm X}$).

Figure~\ref{fig:threeBody} compares the energy of these three bound states, and the excitonic component of each bound state. We again plot the energy as a function of $1/N$, and find it is well fit by a quadratic polynomial, which allows us to extrapolate to $1/N=0$. Here, we find an energy splitting between the $T_+$ bound state and the $S$ or $T_0$ bound states to be $\sim 3$ meV. This splitting can be understood by looking at the exciton wave function in each of the three bound states (panels \textbf{(b)--(d)} in Fig.~\ref{fig:threeBody}). Here we observe that in the case of $T_+$, the exciton can delocalize across both electrons, attracting them closer together without the short-range repulsion due to phase-space filling. This leads to the exciton effectively forming a bonding orbital with the electrons, and thereby lowering the overall energy.

\begin{table}
\centering
\begin{tabular}{>{\centering\arraybackslash}p{3cm} >{\centering\arraybackslash}p{1.5cm} >{\centering\arraybackslash}p{1.5cm}}
\hline\hline
$E^{\uparrow \downarrow}_i -  E^{\uparrow \downarrow}_0$ (meV) & $S$ & $T_0$ \\
\hline
0.00 & 0.29 & 0.04 \\
0.33 & 0.05& 0.27 \\
\hline\hline
\end{tabular}
\caption{Eigenvalues in the spin-unpolarized scenario, and the overlaps (squared) of the eigenstates with the singlet $S$ and triplet $T_0$ initial states. It is seen that the ground state and first excited state predominantly have overlap with the singlet and triplet initial states, respectively.}
\label{Tab:Overlaps}
\end{table}

%\nocite{*}

\bibliographystyle{apsrev4-2}
\bibliography{refs,suppl}% Produces the bibliography via BibTeX.

\end{document}